\documentclass[11pt]{article}
\usepackage{amsmath,amssymb,color,epsfig,cite}
\usepackage{graphicx}
\usepackage{subfigure}
\usepackage{setspace}

\usepackage{amsfonts}

\newcommand{\hoch}[1]{$\, ^{#1}$}

\makeatletter
\@addtoreset{equation}{section}
\makeatother

\newcommand{\be}{\begin{equation}}
\newcommand{\ee}{\end{equation}}
\newcommand{\bea}{\setlength\arraycolsep{2pt} \begin{eqnarray}}
\newcommand{\eea}{\end{eqnarray}}

\def\ft#1#2{{\textstyle{\frac{\scriptstyle #1}{\scriptstyle #2} } }}
\def\fft#1#2{{\frac{#1}{#2}}}

\def\0{{\sst{(0)}}}
\def\1{{\sst{(1)}}}
\def\2{{\sst{(2)}}}
\def\3{{\sst{(3)}}}
\def\4{{\sst{(4)}}}
\def\5{{\sst{(5)}}}
\def\6{{\sst{(6)}}}
\def\7{{\sst{(7)}}}
\def\8{{\sst{(8)}}}
\def\9{{\sst{(9)}}}

\def\sst#1{{\scriptscriptstyle #1}}

\begin{document}



\begin{center}
{\large {\bf Heavy-light Bootstrap from Lorentzian Inversion Formula
}}

\vspace{10pt}
Yue-Zhou Li\hoch{1\dag}

\vspace{15pt}

{\it \hoch{1}Center for Joint Quantum Studies and Department of Physics,\\
School of Science, Tianjin University, Tianjin 300350, China}

\vspace{30pt}

\underline{ABSTRACT}
\end{center}
We study heavy-light four-point function by employing Lorentzian inversion formula, where the conformal dimension of heavy operator is as large as central charge $C_T\rightarrow\infty$. We implement the Lorentzian inversion formula back and forth to reveal the universality of the lowest-twist multi-stress-tensor $T^k$ as well as large spin double-twist operators $[\mathcal{O}_H\mathcal{O}_L]_{n',J'}$. In this way, we also propose an algorithm to bootstrap the heavy-light four-point function by extracting relevant OPE coefficients and anomalous dimensions. By following the algorithm, we exhibit the explicit results in $d=4$ up to the triple-stress-tensor. Moreover, general dimensional heavy-light bootstrap up to the double-stress-tensor is also discussed, and we present an infinite series representation of the lowest-twist double-stress-tensor OPE coefficient. Exact expressions of lowest-twist double-stress-tensor OPE coefficients in $d=6,8,10$ are also obtained as further examples.

\vfill {\footnotesize \hoch{\dag}liyuezhou@tju.edu.cn}

\pagebreak

\tableofcontents
\addtocontents{toc}{\protect\setcounter{tocdepth}{2}}


\newpage
\section{Introduction}
\label{sec:intro}
AdS/CFT correspondence (holography) serves as a bridge connecting gravity theories in anti-de Sitter (AdS) spacetime and strong-coupled CFT living in the AdS boundary \cite{Maldacena:1997re,Gubser:1998bc,Witten:1998qj}, enabling us to exploit conformal field theories (CFT) with sparse spectrum \cite{Heemskerk:2009pn} at strong coupling without referring to any specific CFT theories. On the other hand, although directly studying strongly-coupled CFT is a hard task, recent developments of conformal bootstrap make it achievable. Conformal bootstrap utilizes the conformal symmetry, crossing symmetry, and sometimes other physical consistency conditions such as unitarity to explore the properties of conformal dimensions and operator product expansion (OPE) coefficients in an effective way. In turn, the progress of strongly-coupled CFT can be expected to shed light on some essential aspects of quantum gravity.

In parallel to numerical bootstrap which aims to precisely determine the allowed region of conformal dimensions and OPE coefficients for numerous specific models such as Ising model (see \cite{Poland:2018epd} for a recent review), analytic bootstrap has been developed to probe universality of CFT data in a certain parametric limit. By analyzing the singularities from crossing symmetry near the light-cone limit, the universal spectrum and OPE coefficients of large spin operators were studied extensively, e.g., \cite{Komargodski:2012ek,Fitzpatrick:2012yx,Kaviraj:2015cxa,Alday:2015eya,Kaviraj:2015xsa,Alday:2015ewa}. This progress boosted the large spin perturbation theory \cite{Alday:2016njk}. In particular, the universal data of large spin operators can be asymptotically expanded in terms of the inverse powers of spin $1/J$, and surprisingly, this large spin expansion remains valid even down to finite spin $J$ \cite{Alday:2015ota,Simmons-Duffin:2016wlq}. This incredible validity can be explained by the analyticity in spin in CFT which was made manifest by Caron-Huot Lorentzian inversion formula \cite{Caron-Huot:2017vep,Simmons-Duffin:2017nub,Kravchuk:2018htv}. The Lorentzian inversion formula encapsulates the large spin systematics and allows us to compute OPE coefficients and anomalous dimensions more efficiently, even with finite spin \cite{Cardona:2018dov,Cardona:2018qrt}.

Naturally, Lorentzian inversion formula was applied to investigate quantum gravity and AdS/CFT, for example, it allows us to study correlators up to loop level in supergravity \cite{Alday:2017vkk,Caron-Huot:2018kta} and to understand the growth of extra dimension in AdS/CFT \cite{Alday:2019qrf}. It appears that only pure AdS without any heavy states are considered in the above applications. Undoubtedly, four-point functions with two heavy states, which are referred to as the heavy-light four-point functions $\langle\mathcal{O}_H\mathcal{O}_H\mathcal{O}_L\mathcal{O}_L\rangle$, are interesting and important aspects in CFT as well as in AdS/CFT. In fact, the heavy-light four-point functions are relevant to various topics, e.g. information loss and black hole collapse \cite{Maldacena:2001kr,Fitzpatrick:2014vua,Fitzpatrick:2015zha,Anous:2016kss}, entanglement entropy \cite{Hartman:2013mia,Asplund:2014coa,Chen:2016kyz,Chen:2016dfb,Caputa:2014eta} and chaos \cite{Roberts:2014ifa}, and they are well-studied in AdS$_3$/CFT$_2$ by enjoying the Virasoro symmetry in CFT$_2$. Roughly speaking, it was understood since \cite{Belavin:1984vu} that the Virasoro symmetry completely fixes the Virasoro blocks which contain the contributions of an exchanged operator and its descendants. In particular, at large central charge limit $C_T\rightarrow \infty$, the heavy-light four-point function (the conformal dimension of the heavy operator is $\Delta_H\sim C_T$ and the conformal dimension of the light operator is $\Delta_L\ll C_T$) is sensitive to the Virasoro vacuum block which contains the identity $1$ and all multi-stress-tensors $T^n$. The explicit expression of the Virasoro vacuum block was first computed in \cite{Fitzpatrick:2014vua}. However, the Virasoro symmetry is not available in $d\geq 3$ CFT. It is thus necessary to study heavy-light four-point functions in $d\geq 3$ CFT using different techniques.

Owing to the crossing symmetry, it is simpler to investigate the channel $\mathcal{O}_H\mathcal{O}_L\mathcal{O}_{\Delta,J}\times \mathcal{O}_{\Delta,J}\mathcal{O}_L\mathcal{O}_H$ at first, where the double-twist operators $[\mathcal{O}_H\mathcal{O}_L]_{n,J}$ are exchanged. Holographically, the underlying exchanged operators in this channel were studied recently in \cite{Kulaxizi:2018dxo,Karlsson:2019qfi} by using the bulk phase shift approach \cite{Cornalba:2006xk,Cornalba:2006xm,Cornalba:2007zb,Kulaxizi:2017ixa} or using the Hamiltonian perturbation theory \cite{Kaviraj:2015xsa,Fitzpatrick:2014vua}. In parallel, to search for the universality that is associated with the OPE coefficients of multi-stress-tensor $T^n$ in high dimensions, \cite{Fitzpatrick:2019zqz} proposed a holographic formalism to study the OPE coefficients of multi-stress-tensor, and those OPE coefficients of the lowest-twist sectors exhibit universality by only depending on $\Delta_H,\Delta_L$ and $C_T$. However, the CFT origin of this universality is not clear. By studying stress-tensor commutation relation without holography, \cite{Huang:2019fog} shows that the Virasoso-like structure indeed exists near the light-cone limit. Before long, the OPE coefficient of lowest-twist double-stress-tensor was conjectured in \cite{Kulaxizi:2019tkd} such that the crossing equation is satisfied provided with the holographic results of double-twist operators $[\mathcal{O}_H\mathcal{O}_L]_{n,J}$. It turns out that the OPE coefficients of the lowest-twist double-stress-tensor conjectured in \cite{Kulaxizi:2019tkd} precisely agrees with one found from holography \cite{Fitzpatrick:2019efk}. Recent progress was made in \cite{Karlsson:2019dbd} where the OPE coefficients of the lowest-twist double-stress-tensor, the lowest-twist triple-stress-tensor, and the OPE coefficients and anomalous dimensions of double-twist operators in $d=4$ can all be extracted by solving the crossing equation. Some results in $d=6$ were also obtained up to $T^2$ in \cite{Karlsson:2019dbd}. Remarkably, it can be verified that the data associated with the double-twist operators is consistent with predictions from holography \cite{Kulaxizi:2018dxo,Karlsson:2019qfi}.

The results of \cite{Karlsson:2019dbd} are exciting, but can still be improved, for example, the framework of \cite{Karlsson:2019dbd} relies on their ansatz of heavy-light four-point function near the light-cone limit. Besides, some other questions can be raised. Although the framework of \cite{Karlsson:2019dbd} provides a nice way to compute the relevant OPE coefficients, it could not explain the universality of lowest-twist multi-stress-tensors. Furthermore, the holographic results from bulk phase shift for operators exchanged in $\mathcal{O}_H\mathcal{O}_L\mathcal{O}_{\Delta,J}\times \mathcal{O}_{\Delta,J}\mathcal{O}_L\mathcal{O}_H$ are extracted from eikonal limit (Regge limit \cite{Costa:2012cb}) and they are consistent with the data extracted near light-cone limit \cite{Karlsson:2019dbd}. This fact implies that there is an intersection of the eikonal region and the lowest-twist region. Such a connection between universality in the eikonal region and the lowest-twist region was also discussed in \cite{Fitzpatrick:2019efk}. In this paper, we apply the Lorentzian inversion formula to heavy-light four-point functions back and forth, and we surprisingly find that the Lorentzian inversion formula can shed light on the above questions. Moreover, our results are in precise agreement with those already existed in the literature.

The paper is organized as follows. In section \ref{gene}, we briefly review the conformal blocks and Lorentzian inversion formula. Both the notations used in this paper and preliminary knowledge of heavy-light four-point function are attached in section \ref{heavylight}. We also summarize our main conclusions in section \ref{heavylight}. In section \ref{algo}, we show that heavy-light four-point functions can indeed be bootstrapped by implementing the Lorentzian inversion formula back and forth. In this sense, the resulting CFT data is shown to be universal. We comment on the $\Delta_L$ poles appear in the OPE coefficients of lowest-twist multi-stress-tensor and then we propose an algorithm to manipulate heavy-light bootstrap to extract all universal data. In section \ref{dimfourexam}, we apply our algorithm to work on the examples in $d=4$ up to triple-stress-tensor $T^3$. In section \ref{examgenedim}, we have an attempt at heavy-light bootstrap in general dimension up to double-stress-tensor $T^2$. In particular, an infinite series representation of lowest-twist $T^2$ OPE coefficients is presented. In section \ref{summary}, the paper is summarized and some future directions are discussed. In Appendix \ref{BandBts}, we collect some missing steps of the main text. In Appendix \ref{moreexamsec}, more examples of lowest-twist $T^2$ OPE coefficients are worked out, includes $d=6,8,10$ and a generic pattern.
\section{Generalities}
\label{gene}
In this section, we briefly review the necessary ingredients that will be used throughout this paper, including conformal blocks, Lorentzian inversion formula, and heavy-light four-point function.
\subsection{Conformal blocks}
\label{confblock}
A four-point function $\langle\mathcal{O}_1\mathcal{O}_2\mathcal{O}_3\mathcal{O}_4\rangle$ can be expanded in terms of conformal blocks
\bea
\langle\mathcal{O}_1(0)\mathcal{O}_2(z,\bar{z})\mathcal{O}_3(1)\mathcal{O}_4(\infty)\rangle=\fft{\mathcal{G}(z,\bar{z})}{(z\bar{z})^{\fft{\Delta_1
+\Delta_2}{2}}}\,,\qquad \mathcal{G}(z,\bar{z})=\sum c_{\Delta,J} G^{a,b}_{\Delta,J}(z,\bar{z})\,,
\eea
where $a=(\Delta_2-\Delta_1)/2, b=(\Delta_3-\Delta_4)/2$ and $c_{\Delta,J}$ is the OPE coefficient. The conformal block is the solution of the quadratic Casimir equation
\bea
\mathcal{C}_2 \,G^{a,b}_{\Delta,J}(z,\bar{z})
=\big(\Delta(\Delta-d)+J(J+d-2)\big)G^{a,b}_{\Delta,J}(z,\bar{z})\,,\label{Casi2}
\eea
where
\bea
&& \mathcal{C}_2=\mathcal{D}_z+\mathcal{D}_{\bar{z}}+2(d-2)\fft{z\bar{z}}{z-\bar{z}}
((1-z)\partial_z-(1-\bar{z})\partial_{\bar{z}})\,,
\cr && \mathcal{D}_z=2(z^2(1-z)\partial_z^2-(1+a+b)z^2\partial_z-abz)\,.\label{Casi2full}
\eea
In $d=4$, the closed form of conformal block for scalar four-point function $\langle\mathcal{O}_1\mathcal{O}_2\mathcal{O}_3\mathcal{O}_4\rangle$ is known
\be
G^{a,b}_{\Delta,J}(z,\bar{z})=\fft{z \bar{z}}{z-\bar{z}}(k^{a,b}_{\Delta+J}(z)k^{a,b}_{\Delta-J-2}(\bar{z})-
k^{a,b}_{\Delta+J}(\bar{z})k^{a,b}_{\Delta-J-2}(z))\,,\label{exactblockd4}
\ee
where $k^{a,b}_{\beta}(x)$ is SL(2,R) block and is given by
\be
k^{a,b}_{\beta}(x)=x^{\fft{\beta}{2}}\,_2F_1\Big(a+\fft{\beta}{2},b+\fft{\beta}{2},\beta,x\Big)\,.
\ee
The conformal block (\ref{exactblockd4}) is symmetric under $(z\rightarrow \bar{z},\bar{z}\rightarrow z)$. However, in general dimensions, the exact solutions are hard to come by.

Fortunately, conformal blocks admit series expansion in general dimensions, and the properties of conformal blocks can be analyzed from its series expansion. The colinear expansion around $z\rightarrow 0$ is very useful for our purpose in this paper. The leading term is
\bea
G^{a,b}_{\Delta,J}|_{z\rightarrow 0}= z^{\fft{\Delta-J}{2}} k^{a,b}_{\Delta+J}(\bar{z})\,.\label{leadzex}
\eea
Compare the leading term of conformal block (\ref{leadzex}) (specifying $d=4$) with the exact block in $d=4$ (\ref{exactblockd4}), it is obvious that the terms with $z^{\fft{\Delta-J-2}{2}}$ are missing in the expansion (\ref{leadzex}). (\ref{leadzex}) is referred to as power laws in \cite{Caron-Huot:2017vep}, because it only contains the essential terms with power $z^{(\Delta-J)/2}$. Group theoretically, the full colinear expansion is expected to take the form given by
\bea
G^{a,b}_{\Delta,J}=\sum_{n}\sum_{m=-n}^n B^{a,b}_{n,m}\, z^{\fft{\tau}{2}+n} k^{a,b}_{\beta+2m}(\bar{z})\,,\label{expandz}
\eea
where we denote $\Delta-J=\tau$ and $\Delta+J=\beta$. The coefficients $B^{a,b}_{n,m}$ can be obtained by solving quadratic Casimir equation, see, e.g. \cite{Caron-Huot:2017vep} and Appendix \ref{Bs}.

\subsection{Lorentzian inversion formula}
\label{loren}
Lorentzian inversion formula is a powerful formula to extract the OPE data associated with $s$-channel of four-point function $\langle\mathcal{O}_1\mathcal{O}_2\mathcal{O}_3\mathcal{O}_4\rangle$ \cite{Caron-Huot:2017vep,Simmons-Duffin:2017nub,Kravchuk:2018htv}. The formula is given by
\be
c(\Delta,J)=\fft{1+(-1)^J}{4}\kappa^{a,b}_{\Delta+J}\int dzd\bar{z}\,\mu^{a,b}(z,\bar{z})G^{a,b}_{J+d-1,\Delta-d+1}(z,\bar{z}){\rm dDisc}[\mathcal{G}(z,\bar{z})]\,,\label{Loreninv}
\ee
where $\mu^{a,b}(z,\bar{z})$ is given by
\be
\mu^{a,b}(z,\bar{z})=\Big|\fft{z-\bar{z}}{z\bar{z}}\Big|^{d-2}\fft{\big((1-z)(1-\bar{z})\big)^{a+b}}{(z\bar{z})^2}\,,\label{mu}
\ee
and $\kappa^{a,b}_{\Delta+J}$ is
\be
\kappa^{a,b}_{\beta}=\fft{\Gamma(\fft{\beta}{2}-a)\Gamma(\fft{\beta}{2}+a)\Gamma(\fft{\beta}{2}-b)\Gamma(\fft{\beta}{2}+b)}{2\pi^2 \Gamma(\beta-1)
\Gamma(\beta)}\,.\label{kappa}
\ee
Moreover, ${\rm dDisc}$ represents the double-discontinuity, which is defined by the expectation value of ``squared commutators'', and in practice it is given by
\be
{\rm dDisc}[\mathcal{G}(z,\bar{z})]=\cos(\pi(a+b))\mathcal{G}(z,\bar{z})-\fft{e^{-i(a+b)}}{2}\mathcal{G}^{\circlearrowleft}(z,\bar{z})
-\fft{e^{i(a+b)}}{2}\mathcal{G}^{\circlearrowright}(z,\bar{z})\,,
\ee
where $\mathcal{G}^{\circlearrowleft}$ and $\mathcal{G}^{\circlearrowright}$ are two different analytic continuations for $\bar{z}$ around $1$.
Notice that in Lorentzian inversion formula (\ref{Loreninv}), there is a conformal block with spin and conformal dimension interchanged $G^{a,b}_{J+d-1,\Delta-d+1}$ which is referred to as the funny conformal block (or the inverted conformal block). This funny conformal block is actually related to the light-transform \cite{Kravchuk:2018htv}. Notably, the formula is analytic in spin for $J>1$ except for the factor $(-1)^J$. The factor $(-1)^J$ could be set to $1$ in this paper since exchanged operators can only have even spin. Practically, we should expand $\mathcal{G}(z,\bar{z})$ in terms of cross-channel conformal blocks. Given a certain block with $(\Delta,J)$, we should have
\be
\mathcal{G}(z,\bar{z})=\fft{(z\bar{z})^{\fft{\Delta_1+\Delta_2}{2}}}{\big((1-z)(1-\bar{z})\big)^{\fft{\Delta_2+\Delta_3}{2}}}G^{\tilde{a},\tilde{b}}
_{\Delta,J}(1-\bar{z},1-z)\,,\label{cross-blocks}
\ee
where $\tilde{a}=(\Delta_3-\Delta_2)/2$ and $\tilde{b}=(\Delta_4-\Delta_1)/2$. Then we could perform the inversion integral to obtain $c(\Delta,J)$.

The OPE coefficients are encoded in $c(\Delta,J)$ by \cite{Caron-Huot:2017vep}
\be
c_{\Delta,J}=-{\rm Res}_{\Delta=\Delta'}c(\Delta',J)\,.
\ee
This implies that $c(\Delta',J)$ has poles around physical operators
\be
c(\Delta',J)\sim\fft{c_{\Delta,J}}{\Delta-\Delta'}\,.
\ee
In fact, the integral over $z$ in the Lorentzian inversion formula is responsible for creating the poles above, and the integral over $\bar{z}$ provides other factors that have nothing to do with the poles. To end this subsection, we would like to mention that for the integration over $\bar{z}$, the following formula from \cite{Caron-Huot:2017vep} would be useful throughout our calculation
\bea
&& I^{a,b}_{\hat{\tau}}(\beta)=\int_0^1 \fft{d\bar{z}}{\bar{z}^2}(1-\bar{z})^{a+b}\kappa^{a,b}_\beta k^{a,b}_\beta(\bar{z})\,{\rm dDisc}[
\big(\fft{1-\bar{z}}{\bar{z}}\big)^{\fft{\hat{\tau}}{2}-b}(\bar{z})^{-b}]
\cr &&
\cr && =\fft{\Gamma(\fft{\beta}{2}-a)\Gamma(\fft{\beta}{2}+b)\Gamma(\fft{\beta}{2}-\fft{\hat{\tau}}{2})}
{\Gamma(-\fft{\hat{\tau}}{2}-a)\Gamma(-\fft{\hat{\tau}}{2}+b)
\Gamma(\beta-1)\Gamma(\fft{\beta}{2}+\fft{\hat{\tau}}{2}+1)}\,.\label{intresult}
\eea

\subsection{Heavy-light four-point function}
\label{heavylight}
Our interest is the heavy-light four-point function $\langle\mathcal{O}_H\mathcal{O}_H\mathcal{O}_L\mathcal{O}_L\rangle$ in both $s$-channel and $t$-channel of large central charge $C_T$ CFT ($C_T\sim N^2$) in higher dimension $d>2$, where the conformal dimension of heavy operator is comparable to large $C_T$, i.e. $\Delta_H\sim \mathcal{O}(C_T)$, and the conformal dimension of light operator is $\Delta_L\ll C_T$. To study such a four-point function, we would like to choose a convenient conformal frame in $s$-channel
\be
\langle\mathcal{O}_H(\infty)\mathcal{O}_H(1)\mathcal{O}_L(z,\bar{z})\mathcal{O}_L(0)\rangle\,,\label{framehere}
\ee
where $z,\bar{z}$ are cross ratios. We can then expand (\ref{framehere}) in terms of the conformal blocks of $s$ and $t$-channel to establish the crossing equation. Since we are going to extract the OPE data of both $s$ and $t$-channel, we should clarify the notations used throughout this paper in order to avoid the confusion.

\begin{itemize}
\item[] {\bf Notations}

\item We set the conformal frame (\ref{framehere})  where the crossing equation is
\bea
&& (z\bar{z})^{\Delta_L}\langle \mathcal{O}_H\mathcal{O}_H\mathcal{O}_L\mathcal{O}_L\rangle=\mathcal{G}^s(z,\bar{z})=
\fft{(z\bar{z})^{\Delta_L}}{((1-z)(1-\bar{z}))^{\fft{\Delta_H+\Delta_L}{2}}}\mathcal{G}^t(1-\bar{z},1-z)\,,
\cr &&
\cr && \mathcal{G}^s(z,\bar{z})=\sum c_{\Delta,J}G^{0,0}_{\Delta,J}(z,\bar{z})\,,\qquad \mathcal{G}^t(1-\bar{z},1-z)=\sum
\tilde{c}_{\Delta',J'}G_{\Delta',J'}^{a,b}(1-\bar{z},1-z)\,,
\cr &&\label{notf}
\eea
where $a=(\Delta_2-\Delta_1)/2, b=(\Delta_3-\Delta_4)/2$. When we are using the Lorentzian inversion formula to extract the OPE data of $t$-channel, we simply flip $(z\rightarrow 1-\bar{z},\bar{z}\rightarrow 1-z)$ in above (\ref{notf}).

\item We would denote the $s$-channel as HHLL and the $t$-channel as HLLH.

\end{itemize}

Usually, in large $C_T$ CFT, the OPE coefficients should be expanded in terms of $1/C_T$. A CFT with all data expanded up to the order $\mathcal{O}(1/C_T^0=1)$ is referred to as the generalized free field theory. In generalized free field theory, operators that can be exchanged in HLLH (let us assume $\Delta_H\sim\Delta_L\ll C_T$ for the moment) are double-twist operators $[\mathcal{O}_H\mathcal{O}_L]_{n',J'}$ \cite{Fitzpatrick:2011dm}
\be
[\mathcal{O}_H\mathcal{O}_L]_{n',J'}=\mathcal{O}_H\, \Box^{n'}\partial_{\mu_1}\cdots\partial_{\mu_{J'}} \mathcal{O}_L\,,\qquad \Delta'-J'=\Delta_H+\Delta_L+2n'\,,
\ee
where $n'$ is an integer. For convenience, we use $\tilde{c}_{n',J'}$ to denote the relevant OPE coefficients. An infinite number of double-twist operators accumulate to give rise to the identity of HHLL. The exact free OPE coefficients can be computed by using the Euclidean inversion formula in an elegant way by using the shadow representation, e.g., \cite{Karateev:2018oml}. In fact, the free OPE coefficients are well-known and were obtained in \cite{Fitzpatrick:2011dm}
\bea
\tilde{c}^{\rm free}_{n',J'}&=&\fft{\big(\Delta_H+1-\fft{d}{2}\big)_{n'}\big(\Delta_L+1-\fft{d}{2}\big)_{n'}
\big(\Delta_H\big)_{n'+J'}\big(\Delta_L\big)_{n'+J'}}{n'!J'!\big(\Delta_H+\Delta_L+n'+1-d\big)_{n'}
\big(\Delta_H+\Delta_L+2n'+J'-1\big)_{J'}}
\cr &&
\cr && \times \fft{1}{\big(\Delta_H+\Delta_L+n'+J'-\fft{d}{2}\big)_{n'}\big(J'+\fft{d}{2}\big)_{n'}}\,.\label{freeOPE}
\eea
It behaves like $J'^{\Delta_L-1}$ at heavy-limit and large $J'$ limit. Typically, up to the higher order of large $C_T$ expansion, not only OPE coefficients will be corrected by $1/C_T^n$ with $n\geq1$, but also double-twist operators will acquire anomalous dimensions up to $1/C_T^n$ with $n\geq1$. From the holographic viewpoint, these corrected OPE coefficients and the appeared anomalous dimensions come from tree-level exchange ($n=1$) and loop effects of Witten diagrams ($n>1$). When an additional parametrically large conformal dimension $\Delta_H\sim C_T$ is available in the spectrum, the terms with a higher order of $1/C_T$ have their chance to be compensated by $\Delta_H$. Consequently, the corrections to the OPE coefficients and anomalous dimensions may have contributions up to $\mathcal{O}(1)$ and should not be neglected. Instead, OPE coefficients and anomalous dimensions of the double-twist operators exchanged in HLLH could be expanded in terms of $\Delta_H/C_T$. Follow the convention from \cite{Kulaxizi:2018dxo,Karlsson:2019qfi,Kulaxizi:2019tkd,Karlsson:2019dbd} and for latter convenience, we introduce a parameter $\mu$
\be
\mu=\fft{4\Gamma(d+2)}{(d-1)^2 \Gamma(\fft{d}{2})^2}\fft{\Delta_H}{C_T}\,.
\ee
Naturally, we can organize the double-twist OPE coefficients and anomalous dimensions as follows
\bea
\tilde{c}_{n',J'}(\mu)=\tilde{c}^{\rm free}_{n',J'}\sum_k \mu^k \tilde{c}^{(k)}_{n',J'}\,,\qquad \tilde{\gamma}_{n',J'}(\mu)=\tilde{c}^{\rm free}_{n',J'}\sum_k \mu^k \tilde{\gamma}^{(k)}_{n',J'}\,.\label{orgacg}
\eea
It is worth commenting that the expansion (\ref{orgacg}) is a natural organization: presumably, we can start with full $1/C_T$ expansion and collect those terms having enough power of $\Delta_H$ to reorganize the expansion by arranging $\mu$ order. For the data with $\mathcal{O}(\mu)$ order, $\tilde{c}^{(1)}_{n',J'}$ and $\tilde{\gamma}^{(1)}_{n',J'}$ are contributed by single-stress-tensor exchange in HHLL which is shaped by Ward identity and is proportional to $\mu$, namely
\be
c_{\Delta=d,J=2}=\fft{d^2\Delta_L \Delta_H}{4(d-1)^2C_T}=\mu \fft{\Delta_L \Gamma(\fft{d}{2}+1)^2}{4\Gamma(d+2)}\,.\label{cc1}
\ee
Then $\tilde{c}^{(1)}_{n',J'}$ and $\tilde{\gamma}^{(1)}_{n',J'}$ at large $n'$ and $J'$ could be extracted \cite{Kulaxizi:2018dxo,Karlsson:2019qfi} by using the impact parameter representation at Regge limit \cite{Cornalba:2006xk,Cornalba:2006xm,Cornalba:2007zb,Kulaxizi:2017ixa}. According to the dimensional analysis, $\mathcal{O}(\mu^k)$ corrections to HLLH OPE coefficients and anomalous dimensions are contributed by multi-stress-tensor $T^k$ in HHLL, however, we almost know nothing about $T^k$ OPE coefficients beyond single-stress-tensor. Hence beyond $\mathcal{O}(\mu)$, the expansion (\ref{orgacg}) can only be calculated via holography, either by using bulk phase shift \cite{Kulaxizi:2018dxo,Karlsson:2019qfi} or Hamiltonian perturbation theory \cite{Kulaxizi:2018dxo}. Those holographic investigations are restricted to the limit where OPE coefficients and anomalous dimensions are lying in the large spin regime $\Delta_H\gg J'\gg 1$ (for bulk phase shift approach, the results are actually valid at $n \gg 1$ due to the Regge limit). In addition, the holographic investigations also suggest \cite{Kulaxizi:2018dxo,Karlsson:2019qfi}
\be
\tilde{c}^{(k)}_{n',J'}, \tilde{\gamma}^{(k)}_{n',J'} \sim \fft{1}{J'^{k\fft{d-2}{2}}}\,.\label{scalecg}
\ee
The obtained data at the large spin limit is universal since it turns out that the higher-derivative terms in the gravity theory only contribute to sub-leading large spin. In this paper, we will show the large spin behavior (\ref{orgacg}) is indeed valid by using the Lorentzian inversion formula.

On the other hand, in HHLL, we expect the dominant exchanged operators are multi-stress-tensors $T^k$, for example
\bea
&& k=1\,,\qquad T_{\mu\nu}\,,
\cr && k=2\,,\qquad T_{\mu\nu}\,\Box^{n}\partial_{\mu_1}\cdots\partial_{\mu_{J-4}}T_{\rho\sigma}\,,\cdots
\cr && k=3\,,\qquad T_{\mu\nu}T_{\rho\sigma}\,\Box^{n}\partial_{\mu_1}\cdots\partial_{\mu_{J-6}}T_{\alpha\beta}\,,\cdots
\cr && \cdots\,.
\eea
Analogous to the organization of HLLH data (\ref{scalecg}), the OPE coefficient of $T^k$ could be organized by factorizing $\mu$ out as follows
\be
c_{\Delta,J}=\mu^{k}c^{(k)}_{n,J}\,,\qquad \Delta=kd+J-J_T+2n\,,\qquad J\geq J_T\,,\qquad J_T\leq 2k\,,\label{multistress}
\ee
where $J_T$ is the spin purely contributed by the stress-tensors. However, as we mentioned previously, the  OPE coefficients of multi-stress-tensors are beyond our knowledge, impeding the understanding of $\mathcal{O}(\mu^k)$ corrections to double-twist operators from pure CFT's point of view. The holographic formalism was recently proposed to probe the multi-stress-tensor OPE coefficients in \cite{Fitzpatrick:2019zqz}. By treating heavy operator as a black hole, the heavy-light four-point function can be analyzed as a two-point function on the black hole background, from which the multi-stress-tensor OPE coefficients can be read off. The main conclusion of \cite{Fitzpatrick:2019zqz} is that the lowest-twist multi-stress-tensor OPE coefficients are universal regardless of the detail of the higher derivative gravities they consider. By solving the crossing equation near the light-cone limit provided with the exponentiated ansatz of HHLL correlators that was made in \cite{Karlsson:2019dbd}, \cite{Karlsson:2019dbd} successfully extracted the lowest-twist double-stress-tensor OPE coefficients as well as some low-lying double-twist $[\mathcal{O}_H\mathcal{O}_L]_{n',J'}$ data, where a precise agreement with holographic results \cite{Fitzpatrick:2019efk,Kulaxizi:2018dxo,Karlsson:2019qfi} was observed. However, an insightful CFT understanding of this universality is still lacking. In this paper, we would employ the Lorentzian inversion formula to fill this gap to some extent. Considering that it was observed in \cite{Fitzpatrick:2019efk,Li:2019tpf,Kulaxizi:2019tkd,Fitzpatrick:2019efk,Karlsson:2019dbd} that multi-stress-tensor OPE coefficients have integer $\Delta_L$ poles in even dimension, we will assume $\Delta_L$ is neither an integer nor half-integer (see section \ref{comment}) throughout this paper except for section \ref{comment}. The origin of such poles could be easily observed in our framework, and we will leave the comments in section \ref{comment}.  As guidance for readers, we summarize the main conclusion of this paper below provided with two assumptions

\begin{itemize}
\item[]

{\bf Assumption}:

\item[a.] $\mathcal{O}_L$ belongs to a non-even-integer multiplet: additional light operators with conformal dimension $\tilde{\Delta}_L=\Delta_L+2q$ (where $q$ is an integer) are not available in the spectrum.

\item[b.] $\Delta_L$ is not an integer and half-integer.
\end{itemize}

\begin{itemize}
\item[] {\bf Main conclusion:}

\item[1.] We can bootstrap heavy-light four-point function by implementing the Lorentzian inversion formula back and forth.

\item[2.] The large spin limit of HLLH double-twist data is universal.

\item[3.] The lowest-twist multi-stress-tensor OPE coefficients exchanged in HHLL are universal.

\item[4.] This universality is valid from $z\rightarrow 0, \bar{z}\rightarrow 1$ to $z\rightarrow 0, \bar{z}\rightarrow 0$ with respect to HHLL in the crossing equation (\ref{notf}).
\end{itemize}

\section{Bootstrapping heavy-light: the algorithm}
\label{algo}
In this subsection, we present the generic algorithm for bootstrapping heavy-light four-point functions. By bootstrapping heavy-light, we mean, ambitiously, we would like to have a machine that both details of HHLL and HLLH can come out by following the algorithm. The machine is the Lorentzian inversion formula. The idea is that we would implement the Lorenztian inversion formula from one channel to another channel back and forth to extract all universal CFT data, i.e. $\cdots {\rm HHLL} \rightarrow {\rm HLLH} \rightarrow {\rm HHLL} \cdots$. Typically, the Lorentzian inversion formula is powerful to probe the universality of double-twist operators at large spin limit, elegantly and systematically capturing the large spin perturbation systematics \cite{Kaviraj:2015cxa,Alday:2015eya,Kaviraj:2015xsa,Alday:2015ewa}, in which the asymptotic large spin expansion can be summed over to give rise to the data with finite spin. More surprisingly, in this section, we will show that for heavy-light four-point function where $\Delta_H$ is comparable to $C_T$ charge, the Lorentzian inversion formula provides us the strong evidence that the lowest-twist multi-stress-tensor exchanged in HHLL is universal. In addition, using the Lorentzian inversion formula allows us to have an algorithm computing the lowest-twist multi-stress-tensor OPE coefficients and large spin HLLH double-twist data.

\subsection{Lowest-twist multi-stress-tensor OPE}
\subsubsection{HLLH large spin behavior}
In order to exhibit that Lorentzian inversion formula can encode the multi-stress-tensor data, we would like to start with showing that a certain HHLL block with twist $\tau=\Delta-J$ makes contributions to the OPE and anomalous dimensions of the double-twist operators $[\mathcal{O}_H\mathcal{O}_L]_{n',J'}$ by $1/J'^{\tau/2}$ at the large $J'$ limit. Since we are not restricting ourselves at the leading-twist $n'=0$, we shall keep the expansion $z\rightarrow 0$ up to all order for the HHLL funny conformal block in the Lorentzian inversion formula. In other words, we should use (\ref{expandz}) where we take $\Delta\rightarrow J'+d-1$, $J\rightarrow \Delta'-d+1$. Nevertheless, it is not necessary to know every detail of the expansion (\ref{expandz}), typically, the recursion coefficients $B^{a,b}$ in (\ref{expandz}) actually plays no essential role in the intended parametric limit: it turns out that the recursion coefficients contribute $\mathcal{O}(1)$. Generally, in Lorentzian inversion, we should consider following terms
\be
\fft{\kappa^{a,b}(\beta')}
{\kappa^{a,b}(\beta'+2m)}(1-z)^{a+b}(1-\fft{z}{\bar{z}})^{d-2}G^{a,b}_{J'+d-1,\Delta'-d+1}\big|_{n,m}\sim \tilde{B}^{a,b}_{n,m}\,z^{\fft{J'-\Delta'}{2}+n+d-1}k^{a,b}_{\beta'+2m}(\bar{z})\,,\label{expandfun}
\ee
where $m$ takes integers ranging from $-n$ to $n$, and $\tilde{B}^{a,b}_{n,m}$ is some linear combination of $B^{a,b}$. It turns out the contribution from $\tilde{B}$ is a quantity at the order $\mathcal{O}(1)$ at the heavy and large spin limit, hence $\tilde{B}$ plays no role for the large $J'$ behavior and can be slipped off here for simplicity. On the other hand, the HHLL conformal block with twist $\tau=\Delta-J$ is given by
\be
\mathcal{G}_{HHLL}\sim \fft{(z\bar{z})^{\fft{\Delta_{\rm H}+\Delta_{\rm L}}{2}}}{((1-z)(1-\bar{z}))^{\Delta_{\rm L}}}G^{0,0}_{\Delta,J}(1-\bar{z},1-z)\,.
\ee
To extract large $J'$ limit data, we can take the light-cone limit $\bar{z}\rightarrow 1$ of HLLH, in which $z$ and $\bar{z}$ dependence is factorized. Then we use (\ref{intresult}) to integrate over $\bar{z}$ to yield the following function to be integrated over $z$
\be
C(z,\beta')=z^{\ft{1}{2}(2(n-1)+\Delta_H+\Delta_L-\tau')}\fft{k^{0,0}_{\beta}(1-z)}{(1-z)^{\Delta_L}}I^{(a,a)}_{\tau-\Delta_H-\Delta_L}(\beta'+2m)\,.
\label{C1}
\ee
The $z$ dependence in (\ref{C1}) will not introduce additional $J'$ and $\Delta_H$ dependent factors, and it does nothing but tells us the underlying exchanged operators are double-twist $[\mathcal{O}_H\mathcal{O}_L]_{n',J'}$. Hence, the large $J'$ behavior is encoded in the remaining factor $I^{(a,a)}_{\tau-\Delta_H-\Delta_L}(\beta'+2m)$ lying in the double-twist operator trajectories. For our purpose, we are supposed to take both the heavy and large $J'$ limit. Taking the limit is a little bit subtle here. Precisely we should consider $\Delta_H\gg J'\gg1$. We parameterize $\Delta_H\sim J'/\xi$ and take $\xi\rightarrow 0$ such that we can achieve such a limit and end up with
\be
I^{(a,a)}_{\tau-\Delta_H-\Delta_L}(\beta'+2m)\sim\fft{\Gamma(\Delta_L+J'+m+n)}{\Gamma(-\fft{\tau}{2}+\Delta_L)\Gamma(-\fft{\tau}{2}+J'+m+n+1)}
\rightarrow \fft{J'^{-\ft{\tau}{2}-1+\Delta_L}}{\Gamma(-\fft{\tau}{2}+\Delta_L)}\,.
\ee
Recall that the free OPE coefficients go like $J'^{\Delta_L-1}$, we immediately have
\be
\tilde{c}^{\tau}_{n',J'}\,\,\, {\rm and}\,\,\, \tilde{\gamma}^{\tau}_{n',J'} \sim J'^{-\ft{\tau}{2}}\,,\label{largeJbeh}
\ee
for any twist $n'$, where the superscript $\tau$ denotes that it is contributed by twist $\tau$ conformal block in the cross-channel. However, there is a gap in this rough proof, we skip the large $J'$ behavior of $\tilde{B}^{a,b}_{n,m}$. By solving quadratic Casimir in Appendix \ref{Bts}, we find that for double-twist operators the heavy and large $J'$ limit of $\tilde{B}^{a,b}_{n,m}$ is
\be
\tilde{B}^{a,b}_{n,n}=(-1)^n\fft{\big(\fft{d}{2}-n\big)_n}{\Gamma(n+1)}\,,\qquad \tilde{B}^{a,b}_{n,m<n}=0\,.\label{Bt}
\ee
Thus it does nothing to do with final large $J'$ behavior of HLLH OPE and anomalous dimension.

\subsubsection{Finding lowest-twist multi-stress-tensor}
Next, we would like to show that knowing $\tilde{c}^{k'(d-2)}_{n',J'}$ and $\tilde{\gamma}^{k'(d-2)}_{n',J'}$ with $1\leq k'\leq k$ as HLLH data allows us to find lowest-twist multi-stress-tensor $T^{k+1}$ exchanged in HHLL from Lorentzian inversion formula. The ingredient is the HHLL heavy block. We would like to start with the HLLH heavy block with twist $n'$, which can be deduced from (\ref{expandz}), i.e.,
\be
G^{a,b}_{\Delta',J'}(z,\bar{z})=\sum_{n}\sum_{m=-n}^{m=n} B^{a,b}_{n,m} z^{\fft{1}{2}(2(n'+n) +\Delta_L+\Delta_H+\tilde{\gamma}_{n',J'}(\mu))}\bar{z}^{\ft{\Delta_H+\Delta_L+\tilde{\gamma}_{n',J'}(\mu)}{2}+J'+m+n'}
\,.\label{HLLH-hev}
\ee
where $\Delta'=\Delta_H+\Delta_L+2n'+\tilde{\gamma}_{n',J'}(\mu)$. Crossing (\ref{HLLH-hev}) by taking $(z\rightarrow 1-\bar{z}, \bar{z}\rightarrow 1-z)$ leads to the desired conformal block that will be used to construct $\mathcal{G}$. Note we are restricted to large $J'$ limit where the summation over $J'$ can be replaced by the integration over $J'$, we thus have
\be
\mathcal{G}_{HLLH}=\fft{(z\bar{z})^{\Delta_L}}{((1-z)(1-\bar{z}))^{\fft{\Delta_H+\Delta_L}{2}}}
\sum_{n'}\int_{0}^{\infty} dJ'\tilde{c}_{n',J'}(\mu)\,G^{a,b}
_{\Delta',J'}(1-\bar{z},1-z)\,.\label{GHLLH}
\ee
It is worth noting that (\ref{GHLLH}) is only valid at the limit $z\rightarrow 0$, since HLLH four-point function evaluated at large $J'$ limit by integrating over $J'$ is only consistent with the limit $\bar{z}\rightarrow 1$, namely $z\rightarrow 0$ after crossing. In other words, the large $J'$ data of HLLH evaluated before forces that we can only probe the lowest-twist data in HHLL.

Then as soon as we know $\tilde{c}^{k'(d-2)}_{n',J'}$ and $\tilde{\gamma}^{k'(d-2)}_{n',J'}$ we can know $\mathcal{G}_{HLLH}$ up to the order $\mathcal{O}(\mu^{(k+1)(d-2)})$ by expanding (\ref{HLLH-hev}) in terms of anomalous dimension $\tilde{\gamma}_{n',J'}(\mu)$. Practically, the expansion up to $\mathcal{O}(\mu^{(k+1)(d-2)})$ is permitted, since ${\rm dDisc}$ only keeps terms with $\log^m$ where $m\geq2$, while the unknown information $\tilde{c}^{(k+1)(d-2)}_{n',J'}$ and $\tilde{\gamma}^{(k+1)(d-2)}_{n',J'}$ is attached to linear $\log$ which will always be killed by dDisc. This is analogous to one-loop investigation of supergravity correlator, in which the one-loop effect can be computed by squaring the tree-level data due to the same reason here \cite{Alday:2017vkk,Caron-Huot:2018kta}. At the order $\mathcal{O}(\mu^{(k+1)(d-2)})$, it follows from (\ref{largeJbeh}) that $\tilde{c}^{k'(d-2)}_{n',J'}$ and $\tilde{\gamma}^{k'(d-2)}_{n',J'}$ contributes to the large $J'$ behavior by $J'^{-(k+1)(d-2)/2}$ via many possible combinations, for example,
\be
\tilde{\gamma}^{k(d-2)}_{n',J'}\tilde{c}^{(d-2)}_{n',J'}\,,\qquad \tilde{\gamma}^{(k-1)(d-2)}_{n',J'}\tilde{\gamma}^{2(d-2)}_{n',J'}\,,\qquad \tilde{\gamma}^{(k-1)(d-2)}_{n',J'}\tilde{c}^{(d-2)}_{n',J'}\,,\cdots\,.\label{comb}
\ee
Note as for $\tilde{B}^{a,b}$ in (\ref{expandfun}), $B^{a,b}$ is also of order $\mathcal{O}(1)$ at heavy and large $J'$ limit and hence does not contribute any $J'$ dependence. Precisely, $B^{a,b}$ is given by
\be
B^{a,b}_{n,-n}=\fft{\big(\fft{d}{2}-1\big)_{n}}{\Gamma(n+1)}\,,\qquad B^{a,b}_{n,m>-n}=0\,,\label{B}
\ee
for which the detail is presented in Appendix \ref{Bs}. Then the integration over $J'$ leads to the relevant factor as follows
\be
\mathcal{G}\sim z^{(k+1)(d-2)}\Gamma\Big(\Delta_L-(k+1)(d-2)/2\Big)\,.\label{GHLLHfin}
\ee
All other factors such as $\bar{z}$ dependence, the summation over $n'$, and other $\Delta_L$ dependent coefficients are not relevant for our purpose since the pole that signals the exchanged operators is encoded in $z$ dependence. We keep a Gamma function for later comments in section \ref{comment}. Then Lorentzian inversion formula provided with (\ref{GHLLHfin}) now is
\be
c(\Delta,J)=\int_0^1 dz \,z^{\fft{1}{2}(-2-\tau+(k+1)(d-2))} \mathcal{F}\,,\label{cHHLL}
\ee
where $\mathcal{F}$ is some unknown but regular factors (except for some $\Delta_L$ poles) independent of $z$. It is obvious from (\ref{cHHLL}) that it encodes the OPE coefficients of the lowest-twist multi-stress-tensor $\tau=(k+1)(d-2)$ and we are allowed to compute them by using Lorentzian inversion formula as soon as we know all $\tilde{c}^{k'(d-2)}_{n',J'}$ and $\tilde{\gamma}^{k'(d-2)}_{n',J'}$ with $1\leq k'\leq k$ in HLLH. However, one may worry about the validity of the extracted OPE $c(\Delta,J)$ for low $J$, since Lorentzian inversion formula breaks down at low spin $J<2$ \cite{Caron-Huot:2017vep}. Fortunately, $c(\Delta,J)$ is actually disallowed to have low spin $J$. Recall the conformal dimension of multi-stress-tensors (\ref{multistress}), it is thus clear that the lowest-twist case has $J_T=2k, n=0$, implying $J\geq 2k$.

It is worth noting that one has to be cautious of the procedure discussed in this subsection. Typically, the double-twist operators $[\mathcal{O}_H\mathcal{O}_L]_{n',J'}$ in HLLH are likely to mix with other operators. For example, $[\mathcal{O}_H\mathcal{O}_L]_{n',J'}$ would be mixing with $[\mathcal{O}_H\tilde{\mathcal{O}}_L]_{n'-1,J'}$ where the conformal dimension of $\tilde{\mathcal{O}}_L$ is $\tilde{\Delta}_L=\Delta_L+2$: they share same conformal dimension, twist and spin. In this way, the OPE coefficients $\tilde{c}^{(k)}_{n',J'}$ and anomalous dimensions $\tilde{\gamma}^{(k)}_{n',J'}$ should be interpreted as the weighted average over degenerate operators. Under the weighted average, it is apparent that, e.g. $\langle \tilde{\gamma}^{k(d-2)}_{n',J'}\tilde{c}^{(d-2)}_{n',J'}\rangle$ is not equal to $\langle\tilde{\gamma}^{k(d-2)}_{n',J'}\rangle\langle\tilde{c}^{(d-2)}_{n',J'}\rangle$. Hence the simple combinations (\ref{comb}) are not reliable any more\footnote{We would like to thank Simon Caron-Huot for pointing this out to us.}. Similar mixing problem appears in the efforts on computing loop contribution of supergravity correlators, e.g. \cite{Alday:2017vkk,Caron-Huot:2018kta,Alday:2017xua,Aprile:2018efk}. Therefore, an assumption should be made throughout this paper: there are no other light operators having conformal dimension $\tilde{\Delta}_L=\Delta_L+2q$ where $q$ is an integer. This is assumption $a$ listed in section \ref{heavylight}, we shall call this assumption non-even-integer multiplet assumption.
\subsubsection{The universality}
Now, as assumption $a$ in section \ref{heavylight} is made, we are ready to show the main conclusions of this paper listed in section \ref{heavylight}. The assumption $b$ restricting $\Delta_L$ to non-integer and non-half-integer could actually be quickly observed from the factor of (\ref{GHLLHfin}), we would comment on this assumption in more detail in section \ref{comment} momentarily.

We are ready to analyze the universality associated with heavy-light four-point function. The input is OPE coefficient of single stress-tensor that is completely fixed by Ward identity (\ref{cc1}). For convenience, we present it here again
\be
c_{\Delta=d,J=2}=\fft{d^2\Delta_L \Delta_H}{4(d-1)^2C_T}=\mu \fft{\Delta_L \Gamma(\fft{d}{2}+1)^2}{4\Gamma(d+2)}\,.\label{cc2}
\ee
Remarks are necessary here. This coefficient is exact: it does not require a heavy limit of $\Delta_H$. On the other hand, this coefficient is universal in the sense that it only depends on $\Delta_L$, $\Delta_H$, and $C_T$. Immediately, one can use (\ref{cc2}) to calculate the OPE coefficients and anomalous dimensions of the double-twist operators with large spin limit at the order $\mathcal{O}(\mu)$ via using Lorentzian inversion formula. Since (\ref{cc2}) is universal, and the Lorentzian inversion formula will not introduce additional theory dependent parameters, it immediately follows that $\mathcal{O}(\mu)$ HLLH data at large spin limit are universal. Then as discussed previously, we can keep going: use $\mathcal{O}(\mu)$ HLLH large spin data to extract the OPE coefficients of the lowest-twist double-stress-tensor $T^2$ which are universal because of the universality of $\mathcal{O}(\mu)$ HLLH large spin data. In the next, we could input double-stress-tensor OPE and extract $\mathcal{O}(\mu^2)$ HLLH large spin data. Furthermore, $\mathcal{O}(\mu^2)$ HLLH large spin data could be used to extract triple-stress-tensor $T^3$ OPE. We can employ Lorentzian inversion formula back and forth to do this iteratively, in principle all lowest-twist multi-stress-tensor OPE and large spin double-twist data could be bootstrapped by following the present procedure. Typically, since our input is nothing else but universal data (\ref{cc2}), all the relevant coefficients extracted by going through this procedure, i.e., lowest-twist multi-stress-tensor and large spin double-twist data, are universal. Beyond lowest-twist multi-stress-tensor and large spin limit of double-twist data, our analysis expects no universality. 

It would also be essential to comment on the range of this universality. From our analysis, the universality is guaranteed to be valid for $z\rightarrow 0$ with $\bar{z}$ kept arbitrary with respect to the crossing equation (\ref{notf}). Thus there are no constraints for $\bar{z}$ by our construction. We are allowed to take $\bar{z}\rightarrow 1$ to reach the light-cone limit \footnote{In some literatures, the limit $\bar{z}\rightarrow1, z\rightarrow 0$ is referred to as the double-light-cone limit}. On the other hand, analytically continuing $\bar{z}$ to another sheet and then taking $\bar{z}\rightarrow 0$ (i.e., take $(1-\bar{z}) e^{-2\pi i}\rightarrow 1$) is also permitted: this procedure is expected to give us the correct correlator in the large impact regime of the Regge limit. In this way, we could say this universality holds at both the light-cone limit and the large impact regime of the Regge limit. This explains why the results of double-twist data obtained from bulk phase shift in eikonal or Regge limit are consistent with results extracted near the light-cone limit by taking $J\gg n$ \cite{Karlsson:2019dbd,Fitzpatrick:2019efk}.

\subsection{Comments on $\Delta_L$ poles}
\label{comment}
Before we finally propose the algorithm for bootstrapping heavy-light four-point function, we would like to have a subsection commenting on the $\Delta_L$ poles and explaining why the assumption $b$ in section \ref{heavylight} is necessary. The holographic calculations in even dimensions \cite{Fitzpatrick:2019zqz,Li:2019tpf} implies that the multi-stress-tensor would be suffering from poles $1/(\Delta_L-n)$ where $n$ is an integer. This phenomenon can also be observed from recent CFT investigations \cite{Kulaxizi:2019tkd,Karlsson:2019dbd}. Typically, it shows a pattern, for examples, double-stress-tensor OPE has poles $1/(\Delta_L-2)$ in $d=4$ and $1/((\Delta_L-3)(\Delta_L-4))$ in $d=6$. The origin of these poles is clear in our framework, precisely, it comes from (\ref{GHLLHfin}) as the by-product of the lowest-twist multi-stress-tensor $T^{k+1}$ trajectory. Let us write down the relevant factor here again
\be
P(\Delta_L)=\Gamma\Big(\Delta_L-(k+1)(d-2)/2\Big)\,.\label{poles}
\ee
Now the pattern of such poles is clear:
\begin{itemize}
\item[1.] In even dimensions, all multi-stress-tensor OPE coefficients suffer from integer $\Delta_L$ poles.

\item[2.] In general dimensions, for an even number of stress-tensors, e.g., $T^2, T^4,\cdots$, the corresponding OPE coefficients have some integer poles.

\item[3.] In odd dimensions, for an odd number of stress-tensors, e.g., $T^3, T^5, \cdots$, the corresponding OPE coefficients have some half-integer poles.
\end{itemize}
As discussed in \cite{Fitzpatrick:2019zqz}, the existence of these poles is the result of the fact that HHLL double-twist operators $[\mathcal{O}_L\mathcal{O}_L]_{n,J}$ are not distinguishable from some of the multi-stress-tensor operators for certain $\Delta_L$. Separately, OPE coefficients associated with multi-stress-tensor and double-twist operators in HHLL have same $\Delta_L$ poles. When the value of $\Delta_L$ approaches those poles, relevant multi-stress-tensor and double-twist operators share the same conformal blocks where the divergence in $\Delta_L$ will be identically canceled \cite{Fitzpatrick:2019zqz,Li:2019tpf,Fitzpatrick:2019efk}. Typically, the holographic technique developed in \cite{Fitzpatrick:2019zqz} is not able to read off HHLL double-twist $[\mathcal{O}_L\mathcal{O}_L]_{n,J}$ OPE coefficients. In order to obtain HHLL double-twist data, we are required to relate the data of near-boundary expansion to the data of near-horizon expansion where the near-horizon regularity shall be well-imposed \cite{Fitzpatrick:2019zqz,Li:2019tpf}. Moreover, the holographic techniques are no longer able to determine the mixed OPE coefficients \cite{Fitzpatrick:2019zqz,Li:2019tpf}.

We hope our framework could resolve this situation: we expect that we can distinguishably extract both multi-stress-tensor OPE and HHLL double-twist OPE with poles attached, and clearly observe they merge to eliminate the relevant pole whenever $\Delta_L$ is approaching that pole. Unfortunately, this problem remains unclear till now: Standardly, individual crossed conformal block in $\mathcal{G}$ contributes ( only consider leading-term in the limit $z\rightarrow 0$)
\be
c(\Delta,J)\sim \int z^{\fft{1}{2}(-2-\tau+2\Delta_L)}(\cdots)\,,
\ee
where $\cdots$ represents those $z$-independent factors, resulting in lowest-twist of HHLL double-twist trajectory $[\mathcal{O}_L\mathcal{O}_L]_{n=0,J}$. Thus using the Lorentzian inversion formula without the heavy and large spin limit should standardly lead to the answer of HHLL double-twist OPE coefficients. However, as soon as the heavy limit and large spin limit are both taken, the resulting HLLH correlator would have curious power law of $z$ (\ref{cHHLL}) where HHLL double-twist signals got lost but multi-stress-tensor appears.

Nevertheless, we have to overcome this obstacle for the purpose of going to specific CFT, for examples, $d=4,\mathcal{N}=4$ super-conformal Yang-Mills theory, in which half-BPS operators all have integer conformal dimensions. From the holographic point of view, sphere reductions from type IIB string theory or M theory are more likely to give rise to integer $\Delta_L$ in even dimensions \cite{Bremer:1998zp,Li:2019tpf}. The heavy-light bootstrap with integer or half-integer $\Delta_L$ thus deserves future investigations \cite{Li:2020dqm}. On the other hand, it turns out that when $\Delta_L$ approaches a certain pole, the relevant operators acquire anomalous dimension for which the product of this anomalous dimension and the relevant OPE coefficient could be determined by taking the Residue at that pole of relevant multi-stress-tensor OPE coefficient \cite{Li:2019tpf}. We can also understand, from the viewpoint of the Lorentzian inversion formula, that this anomalous dimension should emerge. Note the relevant term in ${\rm dDisc}$ is $z^{\Delta_L-p}\Gamma(\Delta_L-p)$, where $p$ is the upper bound of involved poles, by expanding around a certain pole $p-p'$, it becomes
\be
z^{\Delta_L-p}\Gamma(\Delta_L-p)\sim \fft{(-1)^{1+p'}z^{p'}}{\Gamma(1+p')}\big(\fft{1}{(p-p'-\Delta_L)}+\log z+\cdots\big)\,,
\ee
where $\cdots$ denotes other irrelevant terms and the divergence term should be expected to be canceled by another set of operators. $\log z$ implies that the corresponding multi-stress-tensor or HHLL double-twist (now they mix with each other) acquire anomalous dimension. We hope our framework could also inspire the understanding of this anomalous dimension and verify the Residue relation proposed in \cite{Li:2019tpf} in the future \cite{Li:2020dqm}.
\subsection{The algorithm}
In this subsection, with assumptions listed in section \ref{heavylight} in hands, we would explicitly propose the algorithm to bootstrap heavy-light four-point function below.
\begin{itemize}
\item[1.] Start with the single-stress-tensor conformal block of HHLL, Lorentzian-invert to extract $\mathcal{O}(\mu)$ HLLH data (OPE coefficients and anomalous dimension of double-twist operators $[\mathcal{O}_{\rm H}\mathcal{O}_{\rm L}]_{n,J}$) in the heavy and large spin limit).
\item[2.] Take advantage of $\mathcal{O}(\mu)$ HLLH data to evaluate $\mathcal{O}(\mu^2)$ colinear ($z\rightarrow 0$) four-point function by summing over twists $n$ and integrating over spin.
\item[3.] Lorentzian-invert $\mathcal{O}(\mu^2)$ colinear four-point function to obtain $\mathcal{O}(\mu^2)$ HHLL OPE data which encodes lowest-twist double-stress-tensor OPE coefficients, read off double-stress-tensor OPE coefficients.
\item[4.] Input lowest-twist double-stress-tensor conformal block of HHLL, Lorentzian-invert to extract $\mathcal{O}(\mu^2)$ HLLH data in the heavy and large spin limit.
\item[5.] Recursively repeat 1 to 4 to extract more and more $\mathcal{O}(\mu^{\rm order})$ HLLH data and lowest-twist $T^{\rm order}$ OPE coefficients of HHLL.
\end{itemize}

\section{Examples in four dimension up to $T^3$}
\label{dimfourexam}
In this section, we follow the algorithm introduced in the previous section to solve the heavy-light four-point function in four dimension up to $T^3$ as an explicit example.
\subsection{$\mathcal{O}(\mu)$ double-twist}
In $d=4$, the closed form of conformal block is known as (\ref{exactblockd4}) which simplifies things a lot. Since the conformal block (\ref{exactblockd4}) is explicitly invariant under interchanging $z$ and $\bar{z}$, making it possible just to use half of it, thus we only need to evaluate
\be
c(\Delta',J')=\int_0^1 dzd\bar{z}\fft{(z-\bar{z})}{(z\bar{z})^3}((1-z)(1-\bar{z}))^{a+b}k^{a,b}_{\beta'}(\bar{z})
k^{a,b}_{2-\tau'}(z){\rm dDisc}[\mathcal{G}_T(z,\bar{z})]\,,\label{examT}
\ee
where $\mathcal{G}(z,\bar{z})$ is single-stress-tensor conformal block, which in $d=4$ is specifically given by (still evaluate a half of  (\ref{exactblockd4}))
\bea
\mathcal{G}_T(z,\bar{z})=-\fft{\Delta_L(z-1)^{-1-\Delta_L}(\bar{z}-1)^{1-\Delta_L}(z\bar{z})^{\ft{\Delta_H+\Delta_L}{2}
}\big(3(1-z^2)+(z^2+4z+1)\log z\big)}{40(z-\bar{z})}\,,\label{dimfourT}
\cr &&
\eea
where the parameter $\mu$ in single-stress-tensor OPE (\ref{cc1}) is slipped off such that we can organize HLLH data exactly following (\ref{orgacg}) in section \ref{heavylight}: we use Lorentzian inversion formula to directly extract $\tilde{c}^{(k)}_{n',J'}$ and $\tilde{\gamma}^{(k)}_{n',J'}$. (\ref{dimfourT}) should be automatically separated into two parts, one is free of $\log$ and one contains $\log z$. The former would be evaluated to contribute to the $\mathcal{O}(\mu)$ corrections to the HLLH double-twist OPE coefficients, and the latter reflects that the HLLH double-twist operators acquire anomalous dimension at the order $\mathcal{O}(\mu)$. Let us first consider the part without $\log$. We can apply the formula (\ref{intresult}) to work out the integration over $\bar{z}$ to obtain the following integral over $z$
\be
\tilde{c}^{(1)}(\Delta',J')=\fft{3\Delta_L}{4} \int dz\, z^{\ft{1}{2}(\Delta_H+\Delta_L-6)}(1-z)^{-\Delta_2}(1+z) k^{a,b}_{4-\tau'}(z)\,.
\ee
To evaluate such an integral, we can expand the hypergeometric function in terms of the series of $z\rightarrow 0$ where each term can be integrated over $z$ and then we can sum over the result of each term to have a nice answer. Taking both the heavy limit $\xi\rightarrow 0$ and large spin limit $J'\rightarrow\infty$ yields
\be
\tilde{c}^{(1)}(\Delta',J')=\fft{3\Delta_L\big(2\Gamma(1-n')\Gamma(1-\Delta_L)+\Gamma(-n')\Gamma(2-\Delta_L)\big)}{
4\Gamma(2-n'-\Delta_L)\Gamma(-1+\Delta_L)}J'^{-2+\Delta_L}\,,
\ee
in which we set $\tau'=\Delta_H+\Delta_L+2n'$. Note the free OPE coefficients (\ref{freeOPE}) with heavy and large spin limit specializing in $d=4$ are
\be
\tilde{c}^{\rm free}_{n',J'}=\fft{\Gamma(\Delta_L+n'-1)}{\Gamma(n'+1)\Gamma(\Delta_L)\Gamma(\Delta_L-1)}
J'^{\Delta_L-1}\,.\label{freeOPEfour}
\ee
Then taking the Residue at a given twist, i.e., integer $n'$ and dividing it by free OPE coefficients (\ref{freeOPEfour}) leads to
\be
\tilde{c}^{(1)}_{n',J'}=-\fft{3\Delta_L(\Delta_L+2n'-1)}{4J'}\,.\label{fourdimc}
\ee
This result exactly agrees with examples of low-lying $n'$ obtained in \cite{Kulaxizi:2018dxo,Karlsson:2019dbd}.

The computation for $\log$ part is similar but more involved. Notably, in previous work on computing anomalous dimension via using the Lorentzian inversion formula, it is not necessary to do the integration over $z$. In most cases, one could just evaluate the integral over $\bar{z}$, and the remaining $z$-dependent integrand will be exactly the same as $z$-dependent integrand associated with OPE data up to an overall $\log z$. Therefore, by definition, the anomalous dimension can be easily worked out by ignoring the $z$-dependent part and projecting everything onto double-twist trajectories. However, in our case, there is a discrepancy between $z$-dependence of $\log$ part and OPE part, which is manifest in (\ref{dimfourT}). The trick here is simply ignoring the overall $\log z$ and integrating the remaining factor over $z$
\be
-\fft{1}{4} \int dz\, z^{\ft{1}{2}(\Delta_H+\Delta_L-4)}(1-z)^{-2-\Delta_2}(1+4z+z^2) k^{a,b}_{4-\tau'}(z)\,.
\ee
This integral does a job to make the double-twist trajectories visible. The limits $\xi\rightarrow0, J'\rightarrow\infty$ should be taken, we thus find
\bea
\tilde{c}^{(1)}_{\log}(\Delta',J')&=&\fft{1}{4\Gamma(2-n'-\Delta_L)\Gamma(\Delta_L-1)\Gamma(\Delta_L)}(\Delta_L(\Delta_L+6n'-1)
\Gamma(-n')\Gamma(1-\Delta_L)\Gamma(\Delta_L)
\cr && -6(-1)^{n'}\Gamma(2-n')\Gamma(2-n'-\Delta_L)
\Gamma(\Delta_L+n'-1))\,.
\eea
Subsequently, we should take the Residue to specify the value at the double-twist trajectories and then divide the resulting expression by free OPE coefficients to end up with anomalous dimension. We end up with the anomalous dimension as follows
\be
\tilde{\gamma}^{(1)}_{n',J'}=-\fft{\Delta_L^2+(6n'-1)\Delta_L+6n'(n'-1)}{2J'}\,.\label{fourdimano}
\ee
It is matching with those examples obtained in \cite{Karlsson:2019dbd}.
\subsection{Lowest-twist double-stress-tensor}
Now we are ready to bootstrap the lowest-twist double-stress-tensor with (\ref{fourdimc}) and (\ref{fourdimano}) in hands. From (\ref{exactblockd4}), the full HLLH block in $d=4$ with bare double-twist operators at the heavy-limit is given by
\be
g_{n',J'}=\fft{(z\bar{z})^{n'+\fft{\Delta_H+\Delta_L}{2}}}{\bar{z}-z}(-z^{J'-1}+\bar{z}^{J'+1})\,.\label{g4}
\ee
As a warm-up exercise, we would present the HLLH four-point function at $\mathcal{O(\mu)}$ order. We would present the individual contribution from the twist $n'$, and then we are supposed to sum over $n'$. For a certain twist $n'$ and $J'$, we have
\be
\mathcal{G}^{HLLH,s,(1)}_{n',J'}(z,\bar{z})=\tilde{c}^{\rm free}_{n',J'} (\tilde{c}^{(1)}_{n',J'}+\fft{\tilde{\gamma}^{(1)}_{n',J'}}{2}(\log z+\log \bar{z}))g_{n',J'}\,,
\ee
where the superscript denotes that it is HLLH at order $\mathcal{O(\mu)}$. Substituting (\ref{freeOPEfour}), (\ref{fourdimc}) and (\ref{fourdimano}) into above, integrating over $J'$ from $0$ to $\infty$ and summing over all twists $n'$ yields (We also need to take $\bar{z}\rightarrow 1$ limit in the end such that the resulting correlator is consistent with large $J'$ limit)
\be
\mathcal{G}^{HLLH,s,(1)}(z,\bar{z})=-\fft{\Delta_L}{4}(1-z)^{-2-\Delta_L}(1-\bar{z})^{1-\Delta_L}(3(1-z^2)+(z^2+4z+1)\log z)(z\bar{z})^{\fft{\Delta_H+
\Delta_L}{2}}\,,
\ee
which is obviously consistent with the HHLL single-stress-tensor block (\ref{dimfourT}). This is the double-check of this approach.

Then we move to the HLLH four-point function at the order $\mathcal{O}(\mu^2)$, specifically, what we are looking at is
\be
\mathcal{G}^{HLLH,s,(2)}_{n',J'}(z,\bar{z})=\fft{\tilde{c}^{\rm free}_{n',J'}}{2} \big(\tilde{c}^{(1)}_{n',J'}\tilde{\gamma}^{(1)}_{n',J'}+\fft{(\tilde{\gamma}^{(1)}_{n',J'})^2}{4}(\log z+\log \bar{z}))(\log z+\log \bar{z}\big)g_{n',J'}\,,\label{GHLLHs2}
\ee
where we ignore the terms contributed by $\tilde{c}^{(2)}_{n',J'}$ and $\tilde{\gamma}^{(2)}_{n',J'}$ since these contributions will be killed by ${\rm dDisc}$. In fact, even $\tilde{c}^{(1)}_{n',J'}$ is useless for the purpose of using Lorentzian inversion formula: it gives us linear $\log$ that becomes trivial under {\rm dDisc}. Integrating over $J'$, summing over $n'$ and turning to cross-channel, we thus have (for simplicity, we only keep $\log^2(1-\bar{z})$ that survives under ${\rm dDisc}$)
\bea
\mathcal{G}^{(2)}_{HLLH}&=&\fft{\Delta_L}{32(\Delta_L-2)}\fft{z^2}{\bar{z}^4}\big(\Delta_L(\Delta_L-1)\bar{z}^4-12\Delta_L(\Delta_L+2)\bar{z}^3
+12(4\Delta_L+3)(\Delta_L+2)\bar{z}^2
\cr && -36(\Delta_L+2)(\Delta_L+1)(2\bar{z}-1)\big)\log^2(1-\bar{z})\,.
\eea
The pole $\Delta_L-2$ in $T^2$ OPE observed in \cite{Fitzpatrick:2019zqz} already appears here. Then we just need to work out the Lorentzian inversion formula (\ref{Loreninv}) with considering the leading $z\rightarrow 0$ term
\be
c(\Delta,J)=-\int dzd\bar{z}\,z^{-\fft{\tau+2}{2}}k^{0,0}_{\beta}(\bar{z})\,{\rm dDisc}[\mathcal{G}^{(2)}_{HLLH}]\,.
\ee
Nevertheless, it is worth noting that we should not apply (\ref{intresult}) anymore, since now no $\bar{z}\rightarrow 1$ limit is assumed. In other words, what we are interested in is finite $J$ result. The following formula would be useful
\be
\int_0^1 d\bar{z}\,\bar{z}^\alpha \,_2F_1(\beta,\beta,2\beta,\bar{z})=\fft{1}{\alpha+1}\,_3F_2(\alpha+1,\beta,\beta;\alpha+2,2\beta;1)\,.
\ee
The trick to do the integral is that we would expand the hypergeometric function in terms of an infinite series which makes the integral doable, and then sum over the infinite series back to an exact result. Meanwhile, the integral over $z$ is not necessarily to be done, since we know it will give rise to the pole $\Delta-J-4$, we only need to slip off $z$ and assign the value $\Delta=J+4$ to the rest. After some algebra, we have
\bea
&& c^{(2)}_{0,J}=\fft{2^{-5-2J}\sqrt{\pi}\,\Delta_L\Gamma(J+1)}{(\Delta_L-2)(J-1)(J-3)(J+6)(J+4)(J+2)\Gamma(J+\fft{3}{2})}
\big(a^{(2)}_0+a^{(2)}_1\Delta_L+a^{(2)}_2\Delta_L^2\big)\,,
\cr &&
\cr && a^{(2)}_0=288\,,\qquad a^{(2)}_1=-(J^4+6J^3-37J^2-138J+72)\,,\qquad  a^{(2)}_2=(J-2)J(J+3)(J+5)\,.\label{TsqOPEfourdim}
\cr &&
\eea
One can straightforwardly verify that (\ref{TsqOPEfourdim}) is exactly same as the holographic result in \cite{Fitzpatrick:2019efk} and also as conjectured in \cite{Kulaxizi:2019tkd}.
\subsection{$\mathcal{O}(\mu^2)$ double-twist and lowest-twist $T^3$}
Going further to work on $\mathcal{O}(\mu^2)$ raises up a practical problem. Typically, there are an infinite number of lowest-twist double-stress-tensors with different spin $J$, and one has to sum over them for the purpose of using Lorentzian inversion formula. This would be a hard-core task, and \cite{Kulaxizi:2019tkd,Karlsson:2019dbd} have done this by taking advantage of a complicated hypergeometric identity. In fact, the summed block exhibits a nice pattern at limit $z \rightarrow 0$ with respect to (\ref{notf}). Based on this nice pattern, \cite{Karlsson:2019dbd} proposed an ansatz to write down all multi-stress-tensor blocks. With the help of that ansatz, \cite{Karlsson:2019dbd} succeeded at obtaining HLLH data and HHLL $T^3$ OPE coefficients that are partly overlapped with this section. The summed lowest-twist double-stress-tensor four-point function is given by (after crossing) \cite{Kulaxizi:2019tkd,Karlsson:2019dbd}
\bea
\mathcal{G}_{T^2}&=&\fft{\Delta_L}{28800(\Delta_L-2)}\big((\Delta_L-4)(\Delta_L-3)(k^{0,0}_{6}(1-z))^2+
\fft{15}{7}(\Delta_L-8)k^{0,0}_{4}(1-z)k^{0,0}_8(1-z)
\cr && +\fft{40}{7}
(\Delta_L+1)k^{0,0}_{2}(1-z)k^{0,0}_{10}(1-z)\big)\,.
\eea
Then exactly as we did in (\ref{examT}) and in previous subsections, we should work out the integral, take the heavy and large spin limit, and then we should take the corresponding Residue. We thus find the corrections to the double-twist OPE coefficients
\bea
\tilde{c}^{(2)}_{n',J'}&=&\fft{1}{96J'^2}\big(27\Delta_L^4+4(27n'-43)\Delta_L^3+3(36n'^2-208n'+39)\Delta_L^2-
4(129n'^2
\cr &&+27n'-7)\Delta_L -624n'(n'-1)\big)\,,\label{fourdimc2}
\eea
and the corrections to the double-twist anomalous dimensions
\bea
\tilde{\gamma}^{(2)}_{n',J'}&=&-\fft{4\Delta_L^3+3(14n'-1)\Delta_L^2+(102n'^2-66n'-1)\Delta_L+34(2n'-1)n'(n'-1)}
{8J'^2}\,,\label{fourdimano2}
\cr &&
\eea
which agree with results obtained by using Hamiltonian perturbation theory \cite{Kulaxizi:2018dxo}. The low-lying examples $n'=0,1,2,3$ of (\ref{fourdimano2}) also exactly match with those obtained in \cite{Karlsson:2019dbd}.

Then we would like to attempt at solving $T^3$ OPE coefficients. Expanding the HLLH heavy block associated with twist $n'$ and spin $J'$ up to $\mathcal{O}(\mu^3)$ leads to (ignoring linear $\log$ term)
\bea
\mathcal{G}^{HLLH,s,(3)}_{n',J'}=\fft{\tilde{c}^{\rm free}_{n',J'}}{8}\Big(\tilde{\gamma}^{(1)}_{n',J'}
(\tilde{c}^{(1)}_{n',J'}+2\tilde{\gamma}^{(2)}_{n',J'})+\fft{1}{6}(\tilde{\gamma}^{(1)}_{n',J'})^3
(\log z+\log \bar{z})\Big)(\log z+\log \bar{z})^2\,.
\cr &&
\eea
By substituting the known data (\ref{fourdimc}), (\ref{fourdimano}), (\ref{fourdimc2}) and (\ref{fourdimano2}) into above, we are allowed to integrate over $J'$ and sum over $n'$ to obtain $\mathcal{G}^{(3)}_{HLLH}$. Although the expression of $\mathcal{G}^{(3)}_{HLLH}$ is too cumbersome and complicated to be presented here, it is for sure that $\log^3(1-\bar{z})$ is involved. After doing the double-discontinuity, we are still left with $\log(1-\bar{z})$. In this way, at the order $T^3$, we have to face with following integral
\be
\int_0^1 d\bar{z}\,\bar{z}^\alpha \,_2F_1(\beta,\beta,2\beta,\bar{z})\log(1-\bar{z})\,.\label{logint}
\ee
Unfortunately, at least to our knowledge, this integral (\ref{logint}) does not have a closed form answer\footnote{We thank Junyu Liu, Wei Li and Jian-Dong Zhang for discussions on this integral.}, while we can only have an infinite series representation for it
\bea
\int_0^1 d\bar{z}\,\bar{z}^\alpha \,_2F_1(\beta,\beta,2\beta,\bar{z})\log(1-\bar{z})=-\sum_{k=0}^{\infty}\fft{2^{2\beta-1}
\Gamma(\beta+\fft{1}{2})^2 \Gamma(k+\beta)^2 (\gamma+\psi(\alpha+k+2))}{\sqrt{\pi}(\alpha+k+1)\Gamma(k+1)\Gamma(\beta)\Gamma(2\beta+k)}\,.
\cr &&
\eea
Thus we are hindered from having lowest-twist $T^3$ OPE coefficients with symbolic $J$ dependence. Nevertheless, for specific $J$, the integral is easy to evaluate, and we could steadily have many low-lying examples for lowest-twist $T^3$ OPE coefficients. We present some examples with low-lying $J=6,8,10,12,14$
\bea
&& c^{(3)}_{0,6}=\frac{\Delta _L (1001 \Delta _L^4+3575 \Delta _L^3+7310 \Delta _L^2+7500 \Delta _L+3024)}{10378368000 (\Delta _L-3) (\Delta _L-2)}\,,
\cr &&
\cr && c^{(3)}_{0,8}=\frac{\Delta _L (3003 \Delta _L^4+6032 \Delta _L^3+9029 \Delta _L^2+7148 \Delta _L+2688)}{613476864000 (\Delta _L-3) (\Delta _L-2)}\,,
\cr &&
\cr && c^{(3)}_{0,10}=\frac{\Delta _L (2431 \Delta _L^4+3077 \Delta _L^3+3742 \Delta _L^2+2216 \Delta _L+888)}{9468531072000 (\Delta _L-3) (\Delta _L-2)}\,,
\cr &&
\cr && c^{(3)}_{0,12}=\frac{\Delta _L (46865039 \Delta _L^4+38644366 \Delta _L^3+41210477 \Delta _L^2+15350374 \Delta _L+8351544)}{3400149507955200000 (\Delta _L-3) (\Delta _L-2)}\,,
\cr &&
\cr && c^{(3)}_{0,14}=\frac{\Delta _L (4892481 \Delta _L^4+2593025 \Delta _L^3+2625560 \Delta _L^2+245300 \Delta _L+477744)}{6497406470370816000 (\Delta _L-3) (\Delta _L-2)}\,.
\eea
The first three examples $J=6,8,10$ are verified to be the same as those in \cite{Karlsson:2019dbd}.

Before ending this section, we would like to comment on what we have learned about the heavy-light bootstrap algorithm from $d=4$ examples. Even though the algorithm is clear and, in principle, it is expected to provide us universal parts of HLLH data and HHLL multi-stress-tensor OPE coefficients up to any high order, some technical issues are impeding our effort on going to higher order. The most important technical issue is that higher order cross-channel four-point functions $\mathcal{G}$ needed in Lorentzian inversion formula requires us to sum over twists $n'$ and spins $J$ for manipulation. In general, higher order calculations come with higher powers of $\log(1-\bar{z})$ in the integral, making the symbolic $J$ formula for $T^n$ OPE coefficients impossible, not mention summing over them. Fortunately, the ansatz of HHLL four-point function proposed in \cite{Karlsson:2019dbd} can release our pressure on summing over all possible $J$ in lowest-twist multi-stress-tensor blocks to pick up required HHLL four-point function $\mathcal{G}_{T^n}$. Typically, $\mathcal{G}_{T^n}$ takes the form of the ansatz proposed in \cite{Karlsson:2019dbd}, where the undetermined coefficients could be fixed by drawing references from some low-lying $J$ OPE coefficients of $T^n$. Thus the HHLL ansatz proposed in \cite{Karlsson:2019dbd} is undoubtedly important for improving our algorithm, which could largely promote efficiency. When it comes to summing over twists $n'$, no difficulty appears in examples $d=4$. However, we will see that this issue is inevitable in the next section. Some other issues exist, and for the moment, we are not aware of the resolution. For example, we will see in the next section that in general dimension, even $\mathcal{O}(\mu)$ order double-twist OPE coefficients can not be solved!

\section{$\mathcal{O}(\mu^2)$ bootstrap in general dimension}
\label{examgenedim}
In this section, we would employ our algorithm to push on $\mathcal{O}(\mu^2)$ heavy-light bootstrap in general dimensions. The main results are as follows:
\begin{itemize}
\item[1.] We find a series representation of $\mathcal{O}(\mu)$ order HLLH double-twist OPE coefficients in general dimensions. Nicely, $\mathcal{O}(\mu)$ order HLLH double-twist anomalous dimension is found with a closed form as $\,_3F_2$ function.
\item[2.] For the lowest-twist double-stress-tensor OPE coefficient in general dimensions, an infinite series representation is provided.
\end{itemize}
\subsection{A warm-up: free double-twist OPE}
As a warm-up, we would like to reproduce the double-twist free OPE coefficients in this subsection. The key ingredient is HLLH funny block in general dimensions, which is an infinite series with relevant terms given by (\ref{expandfun}). For each term, we could take advantage of the nice formula (\ref{intresult}) to integrate over $\bar{z}$ and take the interested limit $\xi\rightarrow 0$ followed by $J'\rightarrow \infty$. We would like to recap the fact that only $\tilde{B}_{n,n}$ survives at heavy-limit as in (\ref{Bt}). Then we find
\bea
c(\Delta',J')|_{n}=\tilde{B}_{n,n}\fft{\Gamma(n-n'+1)\Gamma(1-\Delta_L)}{(n-n')\Gamma(1+n-n'-\Delta_L)\Gamma(\Delta_L)}J'^{\Delta_L-1}\,,
\eea
where we assume $\Delta'-J'=\Delta_H+\Delta_L+2n'$ and $\tilde{B}_{n,n}$ can be found in (\ref{Bt}). We are happy that the summation over $n$ is not hard, we find
\be
c(\Delta',J')=\sum_{n=0}^{\infty}c(\Delta',J')|_n=-\fft{ \Gamma(1-n')\Gamma(\fft{d}{2}-\Delta_L)}{n'\,\Gamma(\fft{d}{2}-n'-\Delta_L)\Gamma(\Delta_L)
}J'^{\Delta_L-1}\,.
\ee
By taking the Residue at integer $n'$, it is straightforward to find
\be
\tilde{c}^{\rm free}_{n',J'}=\fft{(\Delta_L-\fft{d}{2}+1)_{n'}}{\Gamma(n'+1)\Gamma(\Delta_L)}J'^{\Delta_L-1}\,,\label{freed}
\ee
which can be verified to be consistent with heavy and large $J'$ limit of (\ref{freeOPE}). In addition, (\ref{freed}) would come back to (\ref{freeOPEfour}) as soon as $d=4$ is specified.
\subsection{$\mathcal{O}(\mu)$ double-twist}
Now we turn to compute the $\mathcal{O}(\mu)$ correction to HLLH data. The essential ingredient is the form of $\mathcal{G}_T$. Since we are only interested in large $J'$ limit, we could just use the colinear block (\ref{leadzex}) in the cross-channel, we thus have
\be
\mathcal{G}_T=((1-z)(1-\bar{z}))^{\Delta_L}(1-\bar{z})^{\fft{d-2}{2}}(z\bar{z})^{\fft{\Delta_H+\Delta_L}{2}}k^{0,0}_{d+2}(1-z)\,.
\ee
The next step is to address $k^{0,0}_{d+2}(1-z)$. The strategy is to expand the function $k^{0,0}_{d+2}(1-z)$ in terms of an infinite series around $z\rightarrow 0$ where each term can be integrated easily. In the end, we would like to sum the integrated series back to one single expression. Notice that the involved hypergeometric function is $\,_2F_1(\beta,\beta,2\beta,1-z)$ where $\beta=(d+2)/2$, thus we should have following series expansion
\be
\,_2F_1(\beta,\beta,2\beta,1-z)=\sum_{k=0}^{\infty}\fft{\Gamma(2\beta)(\beta)_k^2\big(2(\psi_{k+1}-\psi_{k+\beta})-\log z\big)}{(k!)^2 \Gamma(\beta)^2}z^k\,.
\ee
As expected, we have $\log$ free part and $\log$ part responsible for OPE and anomalous dimensions respectively. Then we would like to obtain anomalous dimensions at first by following the strategy demonstrated in section \ref{dimfourexam}. For each $k$ and $n$ in the heavy and large spin limit, we find
\bea
\tilde{c}^{(1)}_{\log}(\Delta',J')|_{n,k}=\fft{(-1)^{n+1}\Delta_L\Gamma(\fft{d}{2}+k+1)^2 \Gamma(k+n-n')\Gamma(\fft{d}{2}-\Delta_L+2)J'^{\Delta_L-\fft{d}{2}}}
{d^2 \Gamma(\fft{d}{2})\Gamma(k+1)^2 \Gamma(\fft{d}{2}-n)\Gamma(n+1)\Gamma(\fft{d}{2}+k+n-n'-\Delta_L)\Gamma(\Delta_L-\fft{d}{2}+1)}\,.\label{clognk}
\cr &&
\eea
Fortunately, it is not difficult to sum over $n$ and $k$ in (\ref{clognk})
\bea
&&\tilde{c}^{(1)}_{\log}(\Delta',J')=\sum_{n,k=0}^{\infty}\tilde{c}^{(1)}_{\log}(\Delta',J')|_{n,k}
\cr &&
\cr &&=-\fft{\Delta_L \Gamma(-n')\Gamma(d-\Delta_L+1)J'^{\Delta_L-\fft{d}{2}}}{4\Gamma(d-\Delta_L+n'+1)\Gamma(\Delta_L-\fft{d}{2}+1)}\,_3F_2\Big(\fft{d}{2}+1,
\fft{d}{2}+1,-n';1,1,d-n'-\Delta_L+1;1\Big)\,.
\cr &&
\eea
Taking the Residue at double-twist trajectories and dividing the resulting expression by free OPE (\ref{freed}) steadily gives rise to
\be
\tilde{\gamma}^{(1)}_{n',J'}=-\frac{(-1)^{n'} \Gamma (\Delta _L+1) \Gamma (d-\Delta _L+1) \, _3F_2\Big(\frac{d}{2}+1,\frac{d}{2}+1,-n';1,d-n-\Delta _L+1;1\Big)}{2 J'^{\frac{d-2}{2}}\Gamma (d-n-\Delta _L+1) \Gamma (-\frac{d}{2}+n+\Delta _L+1)}\,,\label{anodim}
\ee
which is precisely what \cite{Kulaxizi:2018dxo} obtained by using holographic technique of Hamiltonian perturbation theory.

For $\log$ free part, follow similar analysis, we find
\be
\tilde{c}^{(1)}(\Delta',J')|_{n,k}=-2\tilde{c}^{(1)}_{\log} (\Delta',J')\big(\psi_{k+1}-\psi_{k+(d+2)/2}\big)\,.\label{cnologdimnk}
\ee
The difficulty thus arises. To our knowledge, we can only do the summation over $n$ in (\ref{cnologdimnk}). When it comes to $k$, polygamma functions are involved such that the summation is hard to carry out. Nevertheless, we could take the parametric limit and take the projection onto the double-twist family for each $k$, in which a truncation in the summation over $k$ becomes manifest $k_{\rm max}=n'$. We end up with
\bea
\tilde{c}^{(1)}_{n',J'}\tilde{c}^{\rm free}_{n',J'}=\sum_{k=0}^{n'}\fft{(-1)^{n'-k+1}\Delta_L \Gamma(\fft{d}{2}+k+1)^2 \Gamma(d-\Delta_L+1)\big(\psi_{k+1}-\psi_{k+(d+2)/2}\big)J'^{\Delta_L-d/2}}
{2\Gamma(\fft{d}{2}+1)^2 \Gamma(k+1)\Gamma(n'-k+1)\Gamma(d+k-n'-\Delta_L+1)\Gamma(\Delta_L-\fft{d}{2}+1)}\,.\label{copeseries}
\cr &&
\eea
The simplest case would be the leading-twist $n'=0$, in general dimensions, we have
\be
\tilde{c}^{(1)}_{0,J'}=-\fft{\Gamma(\Delta_L+1)(\gamma+\psi_{(d+2)/2})}{2\Gamma(\Delta_L-\fft{d}{2}+1)}\fft{1}{J'^{\fft{d-2}{2}}}\,.
\ee
When $d$ is even, it is not hard to implement the summation. Particularly, specializing $d=4$ in (\ref{copeseries}) gives back to (\ref{fourdimc}). Some other low-lying examples which are simple enough to present here are $d=6,8$
\bea
&& d=6\,,\qquad \tilde{c}^{(1)}_{n',J'}=-\fft{\Delta_L(60n'(n'+\Delta_L-2))+11(\Delta_L-1)(\Delta_L-2)}{12J'^2}\,,
\cr &&
\cr && d=8\,,\qquad \tilde{c}^{(1)}_{n',J'}=-\fft{5\Delta_L(\Delta_L+2n'-3)\big(5\Delta_L^2+(42n'-15)\Delta_L+2(21
n'^2-63n'+5)\big)}{24J'^3}\,.
\cr &&
\eea
\subsection{An infinite series of lowest-twist $T^2$}
In this section, we would like to see whether we can have access to something on $T^2$ OPE in general dimension. Although we do not even have a closed form for $\mathcal{O}(\mu)$ double-twist OPE coefficients, it is not necessary to include the $\mathcal{O}(\mu)$ double-twist OPE in the correlator as discussed in section \ref{dimfourexam}: they are suppressed by double-discontinuity. Now we need the full heavy-block (\ref{HLLH-hev}) with summing over $n$ in order to implement Lorentzian inversion formula.

Thanks to the heavy and large spin limit such that we have (\ref{B}), we thus find the HLLH four-point function with bare double-twist operators is
\bea
g_{n',J'}(z,\bar{z})=\fft{z^{\fft{\Delta_H+\Delta_L}{2}+n'}\bar{z}^{\fft{\Delta_H+\Delta_L+d}{2}+n'+J'-1}}{(\bar{z}-z)^{\fft{d-2}{2}}}\,,\label{gdim}
\eea
which gives us the relevant term in (\ref{g4}) when specifying $d=4$.\footnote{The reason that another part in (\ref{g4}) is missing in (\ref{gdim}) is: (\ref{gdim}) is deduced from (\ref{expandz}) which is actually the pure power law block where another nonessential power series of $z$ does not exist. On the other hand, (\ref{g4}) is deduced from the full $d=4$ conformal block (\ref{exactblockd4}). In other words, a half of block is enough for our purpose.} Subsequently we will have exactly (\ref{GHLLHs2}) without the contribution from $\tilde{c}^{(1)}_{n',J'}$ (since it is irrelevant). However, we immediately encounter a problem. Following the algorithm, we are required to sum over twists $n'$. Unfortunately, considering that the anomalous dimension in general dimension (\ref{anodim}) is a generalized hypergeometric function and we have no idea how do we simplify such a generalized hypergeometric function, we are not likely to accomplish the summation. Nevertheless, we could keep $n'$ and apply Lorentzian inversion formula to each term with $n'$. Although the involved process is very complicated and it is not appropriate to write all of them down, we manage to have a final answer for lowest-twist $T^2$ OPE contributed by each twist $n'$ by following the standard steps as shown in previous sections. Hence, we end up with an infinite series representation for lowest-twist double-stress OPE coefficients
\bea
&&c^{(2)}_{0,J}=\sum_{n'} \fft{\mathcal{H}(\Delta_L,J)\,_3F_2\Big(\fft{d}{2}+1,\fft{d}{2}+1,-n';1,d-\Delta_L-n'+1;1\Big)^2
}{\Gamma(d-\Delta_L-n'-1)^2\Gamma(\Delta_L-\fft{d}{2}+n'+1)
\Gamma(\Delta_L+J-\fft{d}{2}+n'-1)}\times
\cr &&
\,_3F_2\Big(d+J-2,d+J-2,\Delta_L+J+\fft{d}{2}-2;2(J+d-2),\Delta_L+J+\fft{d}{2}+n'-1;1\Big)\,,
\cr &&
\eea
where $\mathcal{H}(\Delta_L,J)$ is given by
\bea
&& \mathcal{H}(\Delta_L,J)=\ft{16^{2-d-J}\pi^2\Delta_L(d-\Delta_L)(d-\Delta_L-1)\Gamma(\Delta_L+1)\Gamma(J+d-2)^2 \Gamma(d-\Delta_L+1)\Gamma(\Delta_L+J+\fft{d}
{2}-2)}{\Gamma(d+J-\fft{5}{2})\Gamma(d+J-\fft{3}{2})\Gamma(\Delta_L-\fft{d}{2}+1)\sin(\pi \Delta_L)}\,.\label{Ttwodim}
\cr &&
\eea
However, it is rather difficult to start with the infinite series (\ref{Ttwodim}) and try to work out examples with specific dimensions because of the existence of the generalized hypergeometric function. Instead, one should start with anomalous dimensions (\ref{anodim}). We find, for even dimension, (\ref{anodim}) could be reduced to be a nice finite series, making the summation over $n'$ manageable in the process of obtaining $\mathcal{G}^{(2)}_{HLLH}$. After that, the lowest-twist $T^2$ OPE coefficients with symbolic $J$ can be steadily extracted by following the standard integration technique. We present some low-lying examples $d=6,8,10$ in Appendix \ref{moreexamsec}. It should be commented that it seems even dimension is special, while odd dimensions are harder to handle. This is consistent with the holographic treatment of multi-stress-tensor OPE in \cite{Fitzpatrick:2019zqz,Li:2019tpf} where only even dimension case could be truncated to finite series such that the framework is applicable.

\section{Conclusion and future directions}
\label{summary}
In this paper, we studied heavy-light four-point functions by implementing Lorentzian inversion formula back and forth. Focusing on non-degenerate scalar fields and assuming $\Delta_L$ is not an integer and half-integer, we generally show (but not a serious proof) that Lorentzian inversion formula can probe the universality of lowest-twist multi-stress-tensor exchanged in HHLL, and large spin OPE coefficients and anomalous dimensions of double-twist operators exchanged in HLLH. This universality holds within the region $z\rightarrow 0$ with respect to the crossing equation (\ref{notf}). Moreover, an algorithm for computing these data was proposed. In this way, we could state that we can bootstrap heavy-light four-point functions. By applying the algorithm, examples of $d=4$ up to triple-stress-tensor $T^3$ were presented where the results are consistent with results in previous literature. In addition, we also bootstrapped heavy-light four-point function up to $\mathcal{O}(\mu^2)$ ($T^2$) order in general dimensions: we obtain $\mathcal{O}(\mu^2)$ double-twist anomalous dimension in HLLH, series representations of $\mathcal{O}(\mu^2)$ double-twist OPE coefficients in HLLH and series representations of lowest-twist double-stress-tensor OPE coefficients in HHLL.

Although now we can claim that the universality of lowest-twist multi-stress-tensor in heavy-light four-point function is understood by Lorentzian inversion formula to some extent, many related valuable questions are still far from clear. We would like to point out some important future directions
\begin{itemize}
\item The efficiency of our algorithm is somehow limited. \cite{Karlsson:2019dbd} suggests that the first few twists $n'$ of double-twist HLLH data and some low-lying spin $J$ of lowest-twist multi-stress-tensor OPE are enough to maintain the cycle of crossing back and forth and extract more data. It is thus important to investigate the minimum number of twists $n'$ and spin $J$ that is sufficient to maintain the algorithm, which could enhance the efficiency and allow us to go to higher orders.

\item It is clear from Lorentzian inversion formula that lowest-twist multi-stress-tensor OPE coefficients are suffering from some $\Delta_L$ poles. These poles are expected to be canceled by relevant double-twist operators $[\mathcal{O}_L\mathcal{O}_L]_{n,J}$ in HHLL and anomalous dimensions would appear when $\Delta_L$ approaches the poles. Further understanding of this cancelation and inherent anomalous dimensions, alongwith extracting OPE of $[\mathcal{O}_L\mathcal{O}_L]_{n,J}$ is worthy and necessary whenever specific CFTs or supergravities are considered. This understanding, in turn, should shed light on the holographic technique of relating near boundary data to near horizon regularity \cite{Fitzpatrick:2019zqz,Li:2019tpf}.

\item To touch specific CFTs or supergravities, it is also necessary to get rid of the non-even-integer multiplet assumption. It is thus very important and interesting to include other light operators, forming a class of light operators where double-twist operators are mixed. In this situation, there should be extra index such that the double-twist OPE coefficients and anomalous dimensions in HLLH are matrixes, and an appropriate diagonal basis is required.

\item Our results achieve a precise agreement with \cite{Karlsson:2019dbd}, verifying the exponential ansatz in some sense. We wish, similar to Virasoro block in $d=2$, we could somehow directly solve the universal heavy-light conformal block of HHLL which is supposed to be exponentiated. This might be possible by using $6j$ symbol \cite{Liu:2018jhs}.
\end{itemize}
\section*{Acknowledgement}
We are grateful to Simon Caron-Huot for useful discussions. We are also grateful to the JHEP referee for the valuable suggestions, and to Wenliang Li for useful discussion on the modification of the manuscript. This work is supported in part by the NSFC (National Natural Science Foundation of China) Grant No. 11935009 and No.~11875200.

\appendix

\section{Details of $B^{a,b}_{n,m}$ and $\tilde{B}^{a,b}_{n,m}$}
\label{BandBts}
This Appendix is devoted to collect the skipped details in the main text about $B^{a,b}_{n,m}$ and $\tilde{B}^{a,b}_{n,m}$ in the heavy and large spin limit (\ref{B}) and (\ref{Bt}).
\subsection{$B^{a,b}_{n,m}$}
\label{Bs}
At First, we would like to keep track of full $B^{a,b}_{n,m}$ without any limits taken. The logic is simple, we just throw (\ref{expandz}) into quadratic Casimir equation (\ref{Casi2}) and (\ref{Casi2full}), and organize the resulting equation as a recursion equation. We will frequently use two derivative identities for $k^{a,b}_{\beta}(\bar{z})$. The first one is
\bea
\partial^2_{\bar{z}}k^{a,b}_{\beta}=\fft{\big(4ab\bar{z}+\beta(\beta-2)\big)\,k^{a,b}_{\beta}+4
(a+b+1)\bar{z}^2 \partial_{\bar{z}}k^{a,b}_{\beta}}{4\bar{z}^2 (\bar{z}-1)}\,,\label{deiden1}
\eea
which connects second derivative with first derivative without shifting $\beta$. The second identity relates first derivative of $k^{a,b}_\beta$ to $k^{a,b}_\beta$  with $\beta$ shifted by $-2,0,2$, namely
\bea
\partial_{\bar{z}}k^{a,b}_{\beta}&&=\fft{\beta}{2(1-\bar{z})}k^{a,b}_{\beta-2}-\fft{\beta(\beta-2)(a+b)-4ab
}{2(\bar{z}-1)\beta(\beta-2)}k^{a,b}_{\beta}
+\fft{(\beta^2-4a^2)(\beta^2-4b^2)(\beta-2)}{32(\bar{z}-1)(\beta-1)\beta^2(\beta+1)}k^{a,b}_{\beta+2}\,.
\cr &&\label{deiden2}
\eea
Then we do series expansion with respect to $z$ and take advantage of (\ref{deiden1}) and (\ref{deiden2}) such that all derivatives are removed, as results, the Casimir equation becomes
\bea
&& \sum_{m=-n}^{n}(\mathcal{A}_{nm}B^{a,b}_{n,m}k^{a,b}_{\beta+2m}+\mathcal{B}_{n-1}B^{a,b}_{n-1,m}k^{a,b}_{\beta+2m})
 +\sum_{p=1}^n\sum_{m=-n+p}^{p}\fft{1}{\bar{z}^{p-1}}
\big(\fft{1}{z}\,\mathcal{C}^{1,0}_{n-p}k^{a,b}_{\beta+2m}
\cr &&
\cr &&
+\mathcal{C}^{2,-1}_{m}k^{a,b}_{\beta+2(m-1)}+\mathcal{C}^{2,0}_{m}k^{a,b}_{\beta+2m}+
\mathcal{C}^{2,1}_{m}k^{a,b}_{\beta+2(m+1)}\big)B^{a,b}_{n-p,m}=0\,,\label{recurBfull}
\eea
where all $\mathcal{A}, \mathcal{B}, \mathcal{C}$ are given by
\bea
&&\mathcal{A}_{nm}=2\big(m^2+m(\beta-1)+n(n+\tau-d+1)\big)\,, \qquad \mathcal{C}^{1,0}_{n}=-2(d-2)n\,,
\cr &&
\cr && \mathcal{B}_n=-\fft{1}{2}(2a+2n+\tau)(2b+2n+\tau)\,,\qquad \mathcal{C}^{2,-1}_{m}=(d-2)(2m+\beta-\tau)\,,
\cr &&
\cr && \mathcal{C}^{2,0}_{m}=\fft{(d-2)\big(2a(\beta+2m-2)(\beta+2m)+4ab(\tau-2)+(\beta+2m-2)(\beta+2m)(2b+\tau+4n)\big)}
{2(\beta+2m-2)(\beta+2m)}\,,
\cr &&
\cr && \mathcal{C}^{2,1}_m=-\fft{(d-2)\big((2m+\beta)^2-4a^2\big)\big((2m+\beta)^2-4b^2\big)(\beta+\tau+2m-2)}{16
(\beta+2m+1)(\beta+2m)^2 (\beta+2m-1)}\,.
\cr &&
\eea
In addition, another important identity is necessary \cite{Caron-Huot:2017vep}
\be
\fft{k^{a,b}_\beta}{\bar{z}}=k^{a,b}_{\beta-2}+(\fft{1}{2}-\fft{2ab}{\beta(\beta-2)})k^{a,b}_\beta
+\fft{(a^2-\fft{1}{4}\beta^2)(b^2-\fft{1}{4}\beta^2)}{\beta^2(\beta^2-1)}k^{a,b}_{\beta+2}\,.\label{kiden}
\ee
Using this identity (\ref{kiden}) to remove all extra $1/\bar{z}$, the equation (\ref{recurBfull}) boils down to a recursion relation that could be solved for $B^{a,b}_{n,m}$ given boundary condition $B^{a,b}_{0,m}=\delta_{0m}$. Take $n=1$ as examples, we find
\bea
&& B^{a,b}_{1,-1}=\fft{(d-2)(\beta-\tau)}{2(\beta-\tau+d-4)}\,,
\cr &&
\cr && B^{a,b}_{1,0}=\fft{1}{2}\big(a+b+\fft{2ab(4+2\beta-\beta^2+d(\tau-2)-2\tau)}{(\beta-2)
\beta(d-2-\tau)}\big)\,,
\cr &&
\cr && B^{a,b}_{1,1}=\fft{(d-2)(\beta^2-4a^2)(\beta^2-4b^2)(\beta+\tau-2)}{32(\beta-1)
\beta^2 (\beta+1)(\beta+\tau-d+2)}\,.
\eea
The formula (\ref{B}) would come out when we solve $B^{a,b}_{n,m}$ order by order and take the relevant limits. However, this approach is not convincing enough in the sense that we could not find a well-organized closed formula as a solution to the full recursion (\ref{recurBfull}) .

In fact, we can restrict ourselves to bare double-twist trajectories and take the heavy-limit at the very beginning. Surprisingly, as result, the infinite recursion equation would be self-consistently truncated to be a finite and simple one. Taking the heavy-limit reduces (\ref{expandz}) to (\ref{HLLH-hev}) with vanishing $\gamma(\mu)$, i.e.,
\be
G^{a,b}_{\Delta',J'}(z,\bar{z})=\sum_{n}\sum_{m=-n}^{m=n} B^{a,b}_{n,m} z^{\fft{1}{2}(2(n'+n) +\Delta_L+\Delta_H)}\bar{z}^{\ft{\Delta_H+\Delta_L}{2}+J'+m+n'}
\,.\label{HLLH-hevh}
\ee
Subsequently, the quadratic Casimir equation becomes a simple recursion equation
\bea
B^{a,b}_{n,m}&=&-\fft{1}{A^{0,0}_{nm}}(A^{0,-1}_{m-1}B^{a,b}_{n,m-1}+
A^{1,0}_{n-1,m}B^{a,b}_{n-1,m}+A^{1,1}_{n-1,m+1}B^{a,b}_{n-1,m+1}
\cr &&
\cr && +A^{2,1}_{n-2}B^{a,b}_{n-2,m+1})\,,\label{Brecursim}
\eea
where $A$'s are
\bea
&& A^{0,0}_{nm}=-2\big(m^2+m(\beta-1)+n(\tau-d+n+1)\big)\,,\qquad A^{0,-1}_{m}=\fft{1}{2}(2m+2a+\beta)\,,
\cr &&
\cr && A^{1,0}_{nm}=-\fft{1}{2}\big(2(m-n)+\beta-\tau\big)\big(2(m+n+2a-d+2)+\beta+\tau\big)\,,
\cr &&
\cr && A^{1,1}_{nm}=2(m^2+n^2)+2m(\beta-d+1)-(d-2)(\beta-\tau)+2n(\tau-1)\,,
\cr &&
\cr && A^{2,1}_{n}=-\fft{1}{2}(2n+2a+\tau)^2\,.
\eea
We should emphasize that we have already specified $b=a=1/2(\Delta_L-\Delta_H)$ in above recursion (\ref{Brecursim}), and in particular $\tau=\Delta_H+\Delta_L+2n'$ where $n'$ is an arbitrary twist. Then we can take heavy and large spin limit for $A$ in the recursion equation (\ref{Brecursim}). We find for $n>m>-n$
\bea
&&
\fft{A^{1,1}_{n-1,m+1}}{A^{0,0}_{n,m}}=-1\,,\qquad \fft{A^{0,-1}_{m-1}}{A^{0,0}_{n,m}}
=\fft{A^{1,0}_{n-1,m}}{A^{0,0}_{n,m}}=\fft{A^{2,1}_{n-2}}{A^{0,0}_{n,m}}=0\,,\qquad {\rm for}\,\qquad n>m>-n\,.\label{Aratio0}
\cr &&
\eea
For $m=n$, all allowed terms are zero, it is thus clear from (\ref{Aratio0}) that all $B^{a,b}_{n,m>-n}=0$. Then we just need to figure out the recursion equation provided with $m=-n$. Typically, when $m=-n$, only the third term in the right hand side of (\ref{Brecursim}) makes sense, and it is evaluated to be $(-d+4-2n)/(2n)$. Then the recursion equation is largely simplified to be
\be
B^{a,b}_{n,-n}=\fft{d-4+2n}{2n}B^{a,b}_{n-1,1-n}\,,
\ee
which is easy to be solved by
\be
B^{a,b}_{n,-n}=\fft{\big(\fft{d}{2}-1\big)_{n}}{\Gamma(n+1)}\,.
\ee
However, this shall not be the end of story. The reduced block that needs to be solved (\ref{HLLH-hevh}) suffers from ambiguity of $m$. To be precise, for example, relevant $\bar{z}^{m-1}$ in (\ref{HLLH-hevh}) could either be $k^{a,b}_{\beta+2m}/\bar{z}$ or $k^{a,b}_{\beta+2(m-1)}$. Fortunately, this ambiguity is of no significance here, because we could always use (\ref{kiden}) to state $k^{a,b}_{\beta+2m}/\bar{z}$ and $k^{a,b}_{\beta+2(m-1)}$ is equivalent provided with the coefficients in (\ref{kiden}) is vanishing in the heavy-limit. Till now, the proof of (\ref{B}) is completed.

\subsection{$\tilde{B}^{a,b}_{n,m}$}
\label{Bts}
Now we turn to draw (\ref{Bt}) for $\tilde{B}^{a,b}_{n,m}$. We have to remind that this subsection is not a serious proof, but should be served as a strong evidence that (\ref{Bt}) is correct. In fact, as soon as we solve $B^{a,b}_{n,m}$ in (\ref{expandz}) from (\ref{recurBfull}), we could multiply (\ref{expandz}) by the overall factor $\kappa^{a,b}(\beta')/
\kappa^{a,b}(\beta'+2m)(1-z)^{a+b}(1-z/\bar{z})^{d-2}$, then we re-expand it with respect to $z$, organize resulting expansion as (\ref{expandfun}) by using (\ref{kiden}) and turning $(\Delta\rightarrow J+d-1, J\rightarrow \Delta-d+1)$. As the consequence, the coefficients $\tilde{B}^{a,b}_{n,m}$ could be read off \cite{Caron-Huot:2017vep}. Take the heavy and large spin limit, we can observe that (\ref{Bt}) is valid. As in previous subsection on $B^{a,b}_{n,m}$, this approach is not satisfactory since we are not allowed to solve (\ref{Bt}) in an apparent way.

A better way is to take the heavy-limit in the first place. One should note we have a factor $\kappa^{a,b}(\beta')
/\kappa^{a,b}(\beta'+2m)$ attached to each $m$ which is a a little bit annoying and unnatural. For now, we simply do not consider this factor and aim to solve auxiliary coefficients $\hat{B}^{a,b}_{n,m}$ in
\bea
G^{a,b}_{J+d-1,\Delta-d+1}=\sum_n\sum_{m=-n}^n \hat{B}^{a,b}_{n,m} (1-z)^{-a-b}(1-\fft{z}{\bar{z}})^{2-d}z^{-\fft{\tau}{2}+d+n-1}
\bar{z}^{\fft{\beta}{2}+m}\,.
\eea
The resulting recursion equation is infinite but neat
\bea
\hat{B}^{a,b}_{n,m}&=&-\fft{1}{\tilde{\mathcal{A}}^{0,0}_{n,m}}\big(\tilde{\mathcal{A}}^{0,-1}_{m-1}\hat{B}^{a,b}_{n,m-1}
+\tilde{\mathcal{B}}^{1,0}_{n-1,m}\hat{B}^{a,b}_{n-1,m}+\tilde{\mathcal{B}}^{1,-1}_{n-1,m+1}\hat{B}^{a,b}_{n-1,m+1}
\cr && +\sum_{p=2}^n (\tilde{\mathcal{C}}^{1,p}_{n-p,m+p-1}\hat{B}^{a,b}_{n-p,m+p-1}+\tilde{\mathcal{C}}^{2,p}_{n-p,m+p}\hat{B}^{a,b}_{n-p,m+p})
\big)\,,\label{Btrecur}
\eea
where the coefficients are given by
\bea
&& \tilde{\mathcal{A}}^{0,0}_{n,m}=2(m^2+m(\beta-1)+n(n-\tau+d-1))\,,\qquad \tilde{\mathcal{A}}^{0,-1}_{m}=-\fft{1}{2}(2a+2m+\beta)(2b+2m+\beta)\,,
\cr &&
\cr && \tilde{\mathcal{B}}^{1,0}_{n,m}=\fft{1}{2}\Big(4d(m-n)-4n^2+4b(d+n)+2d\beta-4(\beta+2m+2d+b-3)+2(d-b)\tau
\cr && +\tau(4n-\tau)
-4a(b+\tau-n-d+1)\Big)\,,
\cr &&
\cr && \tilde{\mathcal{B}}^{1,-1}_{n,m}=-(d-2)(\beta+\tau+2m-2n-6)\,,\qquad \tilde{\mathcal{C}}^{1,p}_{n,m}=(d-2)\big(\beta+\tau+2(m-n+a+b-2p)
\big)\,,
\cr &&
\cr && \tilde{\mathcal{C}}^{2,p}_{n,m}=-(d-2)\big(\beta+\tau+2(m-n-2p-1)\big)\,.
\eea
Then we take the heavy and large spin limit for these coefficients within double-twist trajectories. For $n>m>-n$, first three terms in the right hand side of (\ref{Btrecur}) tend to zero. Furthermore, for $m=n$, only the first term in the right hand side of (\ref{Btrecur}) makes sense, although it is not zero and actually diverges, it expresses $\hat{B}^{a,b}_{n,n}$ in terms of $\hat{B}^{a,b}_{n,n-1<n}$ which is zero,  indicating that all $\hat{B}^{a,b}_{n,m>-n}=0$. Again we are left with $\hat{B}^{a,b}_{n,-n}$, for which the recursion equation reduces to
\be
\hat{B}^{a,b}_{n,-n}+\fft{d-2}{2n}\sum_{p=1}^{n}\hat{B}^{a,b}_{n-p,-n+p}=0\,,
\ee
which is easily solved by
\be
\hat{B}^{a,b}_{n,-n}=(-1)^n\fft{\big(\fft{d}{2}-n\big)_n}{\Gamma(n+1)}\,.
\ee
Then we would like to recover the factor $\kappa^{a,b}(\beta')
/\kappa^{a,b}(\beta'+2m)$ and translate $\hat{B}$ to $\tilde{B}$. One may naively multiply $\kappa^{a,b}(\beta')
/\kappa^{a,b}(\beta'-2n)$, which, however, identically vanishes in the heavy and large spin limits. This subtlety arises because of the ambiguity of $\bar{z}^m$ exactly the same as in previous subsection. Now we are not lucky enough to make $k^{a,b}_{\beta+2m}/\bar{z}$ and $k^{a,b}_{\beta+2(m-1)}$ equivalent, since the factor $\kappa^{a,b}(\beta')
/\kappa^{a,b}(\beta'+2m)$  is different for each of them. It is possible for us to have the nontrivial result if \footnote{Actually, a more general possibility should be an arbitrary linear combination $\sum_{q=0}^{n} c_{q} k^{a,b}_{\beta-2(n-q)}/\bar{z}^q$ with $\sum_{q}c_q=1$. However, only $c_n$ will come into the final answer while all other $c_i$'s are redundancies. Thus it is natural to shut them down while keep $c_n=1$.}
\bea
G^{a,b}_{J+d-1,\Delta-d+1}|_n=\hat{B}^{a,b}_{n,-n} (1-z)^{-a-b}(1-\fft{z}{\bar{z}})^{2-d}z^{-\fft{\tau}{2}+d+n-1}
\fft{k^{a,b}_{\beta}}{\bar{z}^{n}}\,.
\eea
We then should apply (\ref{kiden}) $n$ times to remove all additional $1/\bar{z}$ factor, and multiplying each term with corresponding $\kappa^{a,b}(\beta')
/\kappa^{a,b}(\beta'+2m)$ factor. Note the factor $\kappa^{a,b}(\beta')
/\kappa^{a,b}(\beta'+2m)$ goes like $\xi^{-2m}$, while coefficients for second and third term in the right hand side of (\ref{kiden}) behave as $\xi$ and $\xi^2$ respectively, we finally find the only surviving term is $\hat{B}^{a,b}_{n,-n}k^{a,b}_{\beta+2n}$, thus
\be
\tilde{B}^{a,b}_{n,n}=\hat{B}^{a,b}_{n,-n}\,,\qquad \tilde{B}^{a,b}_{n,m<n}=0\,,
\ee
which is precisely (\ref{Bt}).
\section{More examples for double-stress-tensor}
\label{moreexamsec}
In this subsection, we present some low-lying examples $d=6,8,10$ for lowest-twist double-stress-tensor OPE coefficients. Actually, from our algorithm of bootstrapping heavy-light four-point function, it is not difficult to work out more even dimensional examples. Typically, we find that lowest-twist double-stress-tensor OPE coefficients in even dimensions follow the pattern as
\bea
&& c^{(2)}_{0,J}=\fft{2^{2-3d-2J}\sqrt{\pi} \Delta_L\Gamma\big(\Delta_L-d+2\big)\Gamma\big(\fft{J-3}{2}\big)\Gamma
\big(\fft{J+d-2}{2}\big)\Gamma(J+d-3)}{\Gamma\big(\Delta_L-\fft{d}{2}+1\big)\Gamma\big(\fft{J+d-3}{2}\big)
\Gamma\big(\fft{J+2d}{2}\big)
\Gamma\big(J+d-\fft{5}{2}\big)}\sum_{i=0}^{\fft{d}{2}}a^{(2)}_i\Delta_L^i \,,
\cr &&
\cr && a^{(2)}_{d/2}=\fft{\Gamma(\fft{J+d-2}{2})\Gamma(\fft{J+2d-1}{2})}{\Gamma(\fft{J-2}{2})\Gamma(\fft{J+d-1}{2})}\,,\qquad a^{(2)}_0={\rm const}\,,\qquad a^{(2)}_{i\neq 0 \wedge d/2}=\sum_{p=0}^{p=d}b^{(2)}_{ip} J^p\,.
\eea
However we do not find patterns governing the constant $a^{(2)}_0$ and other $b^{(2)}_{ip}$. We then just list other $a^{(2)}_i$ or $b^{(2)}_{ip}$ in various dimension below.
\begin{itemize}
\item[] $d=6$
\bea
&& a^{(2)}_0=86400\,,b^{(2)}_{10}=51840\,,b^{(2)}_{11}=45864\,,b^{(2)}_{12}=-1288\,,b^{(2)}_{13}=-1554\,,b^{(2)}_{14}=134
\,,
\cr && b^{(2)}_{15}=42\,,b^{(2)}_{16}=2\,,b^{(2)}_{20}=-8640\,,b^{(2)}_{21}=5796\,,b^{(2)}_{22}=9060\,,
b^{(2)}_{23}=1323\,,b^{(2)}_{24}=-273\,,
\cr && b^{(2)}_{25}=-63\,,b^{(2)}_{26}=-3\,.
\eea

\item[] $d=8$
\bea
&& a^{(2)}_0=67737600\,,b^{(2)}_{10}=82252800\,, b^{(2)}_{11}=24783264\,, b^{(2)}_{12}=-2374984\,, b^{(2)}_{13}=63624\,,
 \cr && b^{(2)}_{14}=120746\,, b^{(2)}_{15}=-9504\,, b^{(2)}_{16}=-3676\,,
 b^{(2)}_{17}=-264\,,  b^{(2)}_{18}=-6\,,  b^{(2)}_{20}=12700800\,,
 \cr && b^{(2)}_{21}=21699216\,,
 b^{(2)}_{22}=4826804\,,   b^{(2)}_{23}=-785444\,,   b^{(2)}_{24}=-171101\,,  b^{(2)}_{25}=26224\,,
 \cr && b^{(2)}_{26}=7006\,,   b^{(2)}_{27}=484\,,  b^{(2)}_{28}=11\,,  b^{(2)}_{30}=-1814400\,,b^{(2)}_{31}=231264\,,b^{(2)}_{32}=1878616\,,
 \cr && b^{(2)}_{33}=710424\,,b^{(2)}_{34}=29146\,,b^{(2)}_{35}=-22704\,,
b^{(2)}_{36}=-4076\,,b^{(2)}_{37}=-264\,,b^{(2)}_{38}=-6\,.
\cr &&
\eea

\item[] $d=10$
\bea
&& a^{(2)}_0=109734912000\,,b^{(2)}_{10}=176795136000\,,b^{(2)}_{11}=29162885760\,,b^{(2)}_{12}=-1932683616
\,,
\cr && b^{(2)}_{13}=245131200\,,b^{(2)}_{14}=-28845960\,,b^{(2)}_{15}=-15354360\,,b^{(2)}_{16}=1926792\,,
b^{(2)}_{17}=615600\,,
\cr && b^{(2)}_{18}=50760\,,b^{(2)}_{19}=1800\,,b^{(2)}_{1,10}=24\,,b^{(2)}_{20}=71937331200\,,b^{(2)}_{21}=41655168000
\,,
\cr && b^{(2)}_{22}=1983391200\,,b^{(2)}_{23}=-723441000\,,
b^{(2)}_{24}=146322800\,,b^{(2)}_{25}=26696250\,,
\cr && b^{(2)}_{26}=-5549250\,,b^{(2)}_{27}=-1363500
\,,b^{(2)}_{28}=-107100\,,b^{(2)}_{29}=-3750\,,b^{(2)}_{2,10}=-50\,,
\cr && b^{(2)}_{30}=4267468800\,,b^{(2)}_{31}=11649074400\,,
b^{(2)}_{32}=4893789960\,,
b^{(2)}_{33}=146590500\,,
\cr && b^{(2)}_{34}=-176081150\,,b^{(2)}_{35}=-9161775\,,b^{(2)}_{36}=5702655
\,,b^{(2)}_{37}=1046250\,,b^{(2)}_{38}=76500\,,
\cr && b^{(2)}_{39}=2625\,,b^{(2)}_{3,10}=35\,,b^{(2)}_{40}=-609638400\,,
b^{(2)}_{41}=-134438400\,,b^{(2)}_{42}=568117440\,,
\cr && b^{(2)}_{43}=332499000\,,b^{(2)}_{44}=53675800\,,
b^{(2)}_{45}=-4186350\,,b^{(2)}_{46}=-2455530\,,b^{(2)}_{47}=-337500\,,
\cr && b^{(2)}_{48}=-22500\,,b^{(2)}_{49}=-750\,,
b^{(2)}_{4,10}=-10\,.
\eea
\end{itemize}
The case $d=6$ was obtained recently in \cite{Karlsson:2019dbd}, which is exactly the same as ours.

\end{document}